\documentclass{aa}

\usepackage{graphicx}
\usepackage{color, colortbl}
\usepackage{orcidlink}
\usepackage{xcolor}
\usepackage{amsmath}
\usepackage{tabularx}
\usepackage{longtable}
\usepackage{dcolumn}
\usepackage{subcaption}
\usepackage{hhline}
\usepackage{txfonts}
\usepackage{hyperref}
\usepackage{cleveref}

\crefname{equation}{Eq.}{Eqs.}
\Crefname{equation}{Equation}{Equations}
\crefname{figure}{Fig.}{Figs.}
\Crefname{figure}{Figure}{Figures}
\crefname{table}{Table}{Tables}
\Crefname{table}{Table}{Tables}
\crefname{chapter}{Chapter}{Chapters}
\Crefname{chapter}{Chapter}{Chapters}
\crefname{section}{Section}{Sections}
\Crefname{section}{Section}{Sections}
\crefname{appendix}{Appendix}{Appendices}
\Crefname{appendix}{Appendix}{Appendices}

\newcommand{\msol}{\,M$_{\odot}$}
\newcommand{\fcbm}{f_{\rm CBM}}
\newcommand{\acbm}{\alpha_{\rm CBM}}
\newcommand{\lb}{\left(}
\newcommand{\rb}{\right)}

\newcommand{\D}[1][]{D_{\text{#1}}}

\newcommand{\Y}[1]{\textbf{Y}^{\text{#1}}}
\newcommand{\Teff}{T_{\rm eff} }

\newcommand\setrow[1]{\gdef\rowmac{#1}#1\ignorespaces}
\newcommand\clearrow{\global\let\rowmac\relax}

\clearrow

\begin{document}

	\title{Probing the physics in the core boundary layers of the double-lined B-type binary KIC\,4930889 from its gravito-inertial modes}

	\author{M. Michielsen \orcidlink{0000-0001-9097-3655}
	\inst{1}
	\and T. Van Reeth \orcidlink{0000-0003-2771-1745}
	\inst{1}
	\and A. Tkachenko \orcidlink{0000-0003-0842-2374}
	\inst{1}
	\and C. Aerts \orcidlink{0000-0003-1822-7126}
	\inst{1,2,3}
}

	\institute{\inst{1}Institute of Astronomy, KU Leuven, Celestijnenlaan 200D, B-3001 Leuven, Belgium\\
		\inst{2}Department of Astrophysics, IMAPP, Radboud University Nijmegen, P.O. Box 9010, 6500 GL Nijmegen, The Netherlands \\
		\inst{3}Max Planck Institute for Astronomy, Koenigstuhl 17, 69117 Heidelberg, Germany \\
		\email{mathias.michielsen@kuleuven.be}}

	\date{Received 07 June 2022; accepted 15 September 2023}

	\titlerunning{Probing the physics in near-core layers}
	\authorrunning{M.\ Michielsen et al.}

	%--------------------------------------------------------------------------------------------
	% \abstract{}{}{}{}{}
	% 5 {} token are mandatory

	\abstract
	% context heading (optional)
	{Stellar evolution models of B-type stars are still uncertain in terms of internal mixing properties, notably in the area between the convective core and the radiative envelope. This impacts age determination of such stars
		in addition to the computation of chemical yields produced at the end of their life.}
	% aims heading (mandatory)
	{ We investigated the thermal and chemical structure and rotation rate in the near-core boundary layer of the double-lined B-type binary KIC\,4930889 from its four-year \textit{Kepler} light curve, ground-based spectroscopy, and \textit{Gaia} astrometry.  }
	% methods heading (mandatory)
	{ We computed grids of 1D stellar structure and evolution models for different mixing profiles and prescriptions of the temperature gradient in the near-core region. We examined the preferred prescription and the near-core rotation rate using 22 prograde dipole modes detected by \textit{Kepler} photometry of KIC\,4930889. We employed a Mahalanobis distance merit function and considered various nested stellar model grids, rewarding goodness of fit but penalising model complexity.}
	% results heading (mandatory)
	{ We were able to constrain the near-core rotation rate of the pulsator to $\Omega_{\text{rot}}=0.73^{+0.02}_{-0.06} \mathrm{d}^{-1}$. Furthermore, we found a preference for either an exponentially decaying mixing profile in the near-core region or absence of additional near-core mixing, but found no preference among the various options for the temperature gradient in this region. The frequency (co)variances of our theoretical predictions are much larger than the errors on the observed frequencies. This forms the main limitation on further constraining the individual parameters of our models. A combination of spectroscopic, astrometric, binary, and asteroseismic information was used to achieve these constraints. Additionally, non-adiabatic pulsation computations of our best models indicate a need for opacity enhancements to accurately reproduce the observed mode excitation.}
	% conclusions heading (optional), leave it empty if necessary
	{ The eccentric close binary system KIC\,4930889 proves to be a promising target to investigate additional physics in close binaries by developing new modelling methods with the capacity to include the effect of tidal interactions for full exploitation of all detected oscillation modes.}


	\maketitle
	%
	%--------------------------------------------------------------------------------------------
	%%%%%%%%%%%%%%%%%%%%%%%%%%%%%%%%%%%%%%%%%%%%%%%%%%%%%%%%%%%%%%%%%%%%%%%%%%%%%%%%%%%%%%%%%%%%%%
	%%%%%%%%%%%%%%%%%%%%%%%%%%%%%%%%%%%%%%%%%%%%%%%%%%%%%%%%%%%%%%%%%%%%%%%%%%%%%%%%%%%%%%%%%%%%%%
	\section{Introduction}

	Slowly pulsating B (SPB) stars are non-radial pulsators with spectral types between B9 and B3, effective temperatures ranging from 11000\,K to 22000\,K, and masses from about 3 to 9\,\msol \,\citep{1991A&A...246..453W}. They are main-sequence stars that exhibit high radial order gravity (g-) mode oscillations, which allows the use of asteroseismology to investigate the physical processes taking place in their interiors. In order to achieve this, space-based monitoring lasting for years is necessary to resolve the frequencies of the g~modes, which have individual periods of roughly 0.5 to 3\,days, with a good enough precision for asteroseismology \citep{Aerts1999,Mathias2001,DeCat2002,Decat2007}. Despite these high demands, the monitoring efforts are worthwhile, as SPB stars have the potential to provide the much needed calibration of the stellar structure and evolution theory for massive stars with convective cores and radiative envelopes. Indeed, the well-known asteroseismic scaling relations used for the low-mass stars based upon their solar-like oscillations cannot be extrapolated to stars with a convective core.
	This leaves poorly calibrated physical properties of their convective core boundary layers, such as its thermal structure or size (often imposed via a free parameter).

	Gravity-mode asteroseismology of main-sequence stars saw its birth thanks to the five-month light curves assembled by the CoRoT space telescope \citep{Degroote2010,Neiner2012} and underwent a major boost from the four-year datasets from the \textit{Kepler} mission \citep[e.g.][for a review]{Aerts2021}. It has meanwhile been applied to numerous cases, both in SPB stars and less massive $\gamma$ Doradus stars to probe a variety of physical processes such as near-core rotation rates \citep[e.g.][]{2016A&A...593A.120V,2020MNRAS.491.3586L,2020A&A...635A.106T}, chemical mixing in the radiative envelope \citep[e.g.][]{2020ApJ...895...51M,2022ApJ...925..154M}, magnetic fields \citep[e.g.][]{2018A&A...616A.148B,2019A&A...627A..64P,2022MNRAS.512L..16L}, opacities \citep[e.g.][]{2018MNRAS.478.2243S,2019MNRAS.485.3544W}, and convective boundary mixing (CBM) whether looking solely at chemical element transport \citep[e.g.][]{2016ApJ...823..130M} or also altering the thermal structure in this CBM region \citep[e.g.][]{2021NatAs...5..715P,2021A&A...650A.175M}.
	These modelling efforts typically involve a parameter study in a multidimensional space, which is generally computationally intense and time-consuming. In order to be able to perform asteroseismic modelling of a sample of SPB stars covering the entire core-hydrogen burning stage and rotational frequencies from almost zero up to the critical rate within a reasonable computation time, \citet{2021NatAs...5..715P} developed a statistical approach to get an initial estimate for the parameters and internal properties of 26 sample SPB stars.

	Our work concerns a novel methodological framework, which we applied to one of these 26 sample stars, namely the double-lined spectroscopic binary SPB KIC\,4930889. In order to achieve this, we relied on the most recent spectroscopic analyses and frequency determinations done since the study by
	\citet{2021NatAs...5..715P}.
	The emphasis of this work is on the CBM processes including the thermal structure and the near-core rotation rate of this binary g-mode pulsator. We first provide an overview of its observed properties in the next section.

	%%%%%%%%%%%%%%%%%%%%%%%%%%%%%%%%%%%%%%%%%%%%%%%%%%%%%%%%%%%%%%%%%%%%%%%%%%%%%%%%%%%%%%%%%%%%%%
	\section{Gravito-inertial pulsations in the double-lined spectroscopic binary KIC\texorpdfstring{\,}{}4930889}

	\citet{2017A&A...598A..74P} analysed the \textit{Kepler} light curve of the system and identified a g-mode period series consisting of 20 pulsations with mode identification $(\ell,m) = (1,1)$ and consecutive radial orders, where a positive $m$-value denotes prograde modes. Additionally they gathered high-resolution spectra of the target using the HERMES spectrograph \citep{2011A&A...526A..69R}. Their spectral synthesis of the disentangled spectra of the system resulted in the parameters listed in \cref{tab:spectro_parameters_pap}. The analysis places both components in the SPB instability strip, which raises the question as to which star the prograde dipole mode pattern belongs.

	\begin{table}[ht]
		\caption{Parameters obtained from the spectroscopic analysis by \citet{2017A&A...598A..74P}.}
		\label{tab:spectro_parameters_pap}
		\centering
		\begin{tabular}{l c c }
			\hline
			\hline
			Parameter & KIC\,4930889\,A & KIC\,4930889\,B  \\
			\hline
			$\Teff$ [K] & 15100 $\pm$ 100 & 12070 $\pm$ 200 \\
			$\log\,g$ [dex]  & 3.95  $\pm$ 0.1 & 3.85  $\pm$ 0.1 \\
			$[$M/H$]$ & $-0.08  \pm 0.1$ & $-0.09  \pm 0.1$ \\
			$v \sin i \,[\text{km s}^{-1}] $ & 116 $\pm$ 6 & 85 $\pm$ 5 \\
			$\xi_{\rm t}$[km$\,$s$^{-1}$] & 1.85 $\pm$ 0.8 & 2 $\pm$ 1 \\
			Light factor  & 0.71 $\pm$ 0.01 & 0.29 $\pm$ 0.01 \\
			Spectral type & B5 IV-V & B8 IV-V \\
			\hline
			Orbital period [d]& \multicolumn{2}{c}{18.296 $\pm$ 0.002} \\			
			q [M$_2$/M$_1$]& \multicolumn{2}{c}{0.77$\pm$0.09} \\
			e              & \multicolumn{2}{c}{0.32$\pm$0.02} \\
			$\omega (^{\circ})$ & \multicolumn{2}{c}{352.7$\pm$4.9} \\
			 $a \sin i_{\rm orb} (R_{\odot}$) & 23.9 $\pm$ 0.5 & 31.1  $\pm$ 0.5 \\
			\hline
		\end{tabular}
		\begin{flushleft}
		\textbf{Notes.} The quoted errors are $1\sigma$ statistical uncertainties, not taking into account any systematic uncertainties.
		\end{flushleft}
	\end{table}

	Asteroseismic modelling, assuming that the primary is the pulsator and based on a statistical approximation for both stellar model properties and mode frequencies, was performed by \citet{2021NatAs...5..715P}. The parameters of their best stellar models are listed in \cref{tab:natastron_param}. While such an approximative statistical approach was chosen to keep simultaneous treatment of 26 pulsators within reasonable computation time, it can never be as detailed as an approach tuned to a particular star, as offered here.
	Moreover, the results listed in  \cref{tab:natastron_param}
	were based on the frequency list and spectroscopic parameters derived by \citet{2017A&A...598A..74P}.
	\citet{2019MNRAS.482.1231J} revisited and renormalised the spectra obtained by \citet{2017A&A...598A..74P}, deriving a new spectroscopic solution listed in \cref{tab:spectro_parameters_jon}. They found their solution to be in better agreement with the evolutionary expectations of their isochrone-cloud modelling. We rely on these spectroscopic parameters to guide the modelling.

	\begin{table}[ht]
	\caption{Stellar parameters of the best asteroseismic models of KIC\,4930889 derived by \citet{2021NatAs...5..715P}.}
	\label{tab:natastron_param}
	\centering
	\begin{tabular}{l c c c c}
		\hline
		\hline
		Parameter & \multicolumn{4}{c}{KIC\,4930889\,A}  \\
		& $\psi_1$ & $\psi_2$ & $\psi_6$ & Average ($\psi_1$, $\psi_5$) \\
		\hline
		$M_{\rm ini}$ [\msol] & 4.375 & 4.0 & 4.2 & 4.06$\pm$0.31  \\
		$Z_{\rm ini}$ & 0.0092 & 0.01  & 0.014 & 0.00924 $\pm$ 0.00002\\
		$\fcbm$  & 0.0128 & 0.036 & 0.03 & 0.012$\pm$0.001 \\
		log($\D[env]$) & 2.736 & 4.6 & 5 & 3.3$\pm$0.5 \\
		$X_{\rm c}/X_{\rm ini}$ & 0.37 & 0.50 & 0.55 & 0.362$\pm$0.0007\\
		$\Omega_{\rm rot}$ (d$^{-1}$) & & & & 0.740$\pm$0.008 \\
		\hline
	\end{tabular}

	\begin{flushleft}
		\textbf{Notes.} From top to bottom, the parameters are  initial mass, metallicity, exponential CBM parameter, envelope mixing at the CBM interface, and central hydrogen content.  The columns list the best stellar models from different grids: their best grid $\psi_1$ which has constant envelope mixing, and grids $\psi_2$ and $\psi_6$ which have the same prescription for envelope mixing as employed in this work. The average and errors listed are the updated values from \citet{2022ApJ...930...94P} for their indistinguishable grids. The rotation frequency is from \citet{2022ApJ...940...49P}.
	\end{flushleft}
\end{table}

\begin{table}[ht]
	\caption{Updated parameters obtained from the spectroscopic analysis by \citet{2019MNRAS.482.1231J}.}
	\label{tab:spectro_parameters_jon}
	\centering
	\begin{tabular}{l c c }
		\hline
		\hline
		Parameter & KIC\,4930889\,A & KIC\,4930889\,B  \\
		\hline
		$\Teff$ [K] & 14020 $\pm$ 280 & 12820 $\pm$ 900 \\
		$\log\,g$ [dex]  & 3.55  $\pm$ 0.24 & 4.38  $\pm$ 0.10 \\
		\hline
	\end{tabular}
	\begin{flushleft}
	\textbf{Notes.} The quoted errors are $1\sigma$ statistical uncertainties, not taking into account any systematic uncertainties.
	\end{flushleft}
\end{table}

	To compute the stellar luminosity, we used

	\begin{equation}
		\log \frac{ L }{ L_{\odot} } = -0.4(M_{\rm S_{\lambda}} + BC_{\rm S_{\lambda}} - M_{\rm bol,\odot})
	\end{equation}
	with
	\begin{equation}
		M_{\rm S_{\lambda}} = m_{\rm S_{\lambda}} - 5\log \frac{d}{10 \rm pc} - R_{\rm S_{\lambda}} E(B-V).
	\end{equation}
	 In this expression, $m_{\rm S_{\lambda}}$ represents the apparent magnitude from the \textit{Gaia} \textit{G}-band, and $d$ is the \textit{Gaia} eDR3 distance from \citet{2021AJ....161..147B}. The $E(B-V)$ reddening value was obtained with the 3D reddening map (Bayestar19) from \citet{2019ApJ...887...93G}, and the reddening vector $R_{\rm S_{\lambda}}$ equals 3.002 for the \textit{G} band \citep{2020MNRAS.495.2738P}. The bolometric correction $BC_{\rm S_{\lambda}}$ was calculated adopting the prescription from \citet{2020MNRAS.495.2738P} (model 3, LTE+non-LTE), and $ M_{\rm bol,\odot}=4.74$.

	However, the system is a close binary, yet it was treated as a single object in \textit{Gaia} DR3. To derive the luminosity for both components separately, we therefore had to adjust their apparent magnitudes based on their respective light factors (here denoted as $f_i$). Given that the apparent magnitude of the system depends on the total observed flux divided by the zero-point (for simplicity denoted as $F$)

	\begin{equation}
		m_{\rm S_{\lambda}, system} = -2.5\log(F),
	\end{equation}
	we can calculate the apparent magnitude of the individual components by adjusting the total apparent magnitude based on their respective light factors

	\begin{align}
		m_{\rm S_{\lambda}, component_{i}} = -2.5\log(f_i \cdot F) &= -2.5\log(F) -2.5\log(f_i) \\ &= m_{\rm S_{\lambda}, system} -2.5\log(f_i).
	\end{align}
	To derive the light factors for the spectroscopic solution from \citet{2019MNRAS.482.1231J}, we followed Eq.\,3 from \citet{2015A&A...581A.129T}. We calculated the ratio of continuum intensities for atmospheric models with parameters given in \cref{tab:spectro_parameters_jon} weighed by the \textit{Gaia} eDR3 \textit{G} passband transmissivity, and used the radii ratio of $\mathcal{R}=R_{secondary}/R_{primary}=0.76$ from \citet{2019MNRAS.482.1231J}. This lead us to light factors of $0.67\pm0.01$ and $0.33\pm0.01$ for the primary and secondary, respectively. Using these corrections to the apparent magnitude, we derived a luminosity $\log \frac{ L }{ L_{\odot} }=2.61\pm0.04$ for the primary, and $\log \frac{ L }{ L_{\odot} }=2.23\pm0.08$ for the secondary.

	Both components of the binary fall within the SPB instability strip, and they both contribute a significant amount to the total observed flux of the system.
	To aid the determination of the pulsator, we divided the \textit{Kepler} light curve in 20 segments according to orbital phase, using the methodology from \citet{2023A&A...671A.121V}, in search of pulsation amplitude and phase modulations as a function of the binary orbital phase. In close binaries, these can be caused by either tidal perturbations \citep[e.g.][]{Reyniers2003a,Reyniers2003b,SamadiGhadim2018,Bowman2019,Steindl2021} or tidal tilting \citep[e.g.][]{SpringerShaviv2013,Handler2020,Kurtz2020,Fuller2020}. In intermediate to wide binaries, these can be caused by the light travel time effect \citep[e.g.][]{2012MNRAS.422..738S,2014MNRAS.441.2515M}. Only one dominant mode turned out to be part of a multiplet separated by the orbital frequency in the Fourier transform of the star's light curve. This mode is f$_{43}$ from \cref{tab:complete_freq_list} ($1.1949135\pm0.0000019$ d$^{-1}$), and the multiplet consists of only two modes. Moreover, none of the detected pulsation phase and amplitude modulations are consistent with each other, except when the modulation is zero within the uncertainty.
	 This is contrary to what is expected for frequency modulations caused by binarity \citep{2012MNRAS.422..738S}. The tidal perturbation of pulsations can affect individual modes more strongly than others, depending on their geometry. However, for g-modes that form a period-spacing pattern (i.e. have the same geometry) we expect the orbital multiplet structure to be very similar for all modes, and most easily detectable for the dominant g-modes \citep{2023A&A...671A.121V}. This was also not the case.
	From this we conclude that neither tidal perturbations or tilting nor the light travel time effect can be detected for this target.

	Additionally, we attempted to determine the pulsator via line profile variations in the 26 spectra observed by \citet{2017A&A...598A..74P}. However, due to the relevant lines being either very weak or broad and shallow, the S/N of the spectra proved insufficient to draw any conclusions from this approach. We therefore make no assumption regarding which component hosts the pulsations, in contrast to the previous study by \citet{2021NatAs...5..715P} who assumed the primary to be the pulsator. We subsequently show that this is the least likely of the two scenarios.

\begin{figure}[t!]
		\centering
		\includegraphics[width=\hsize]{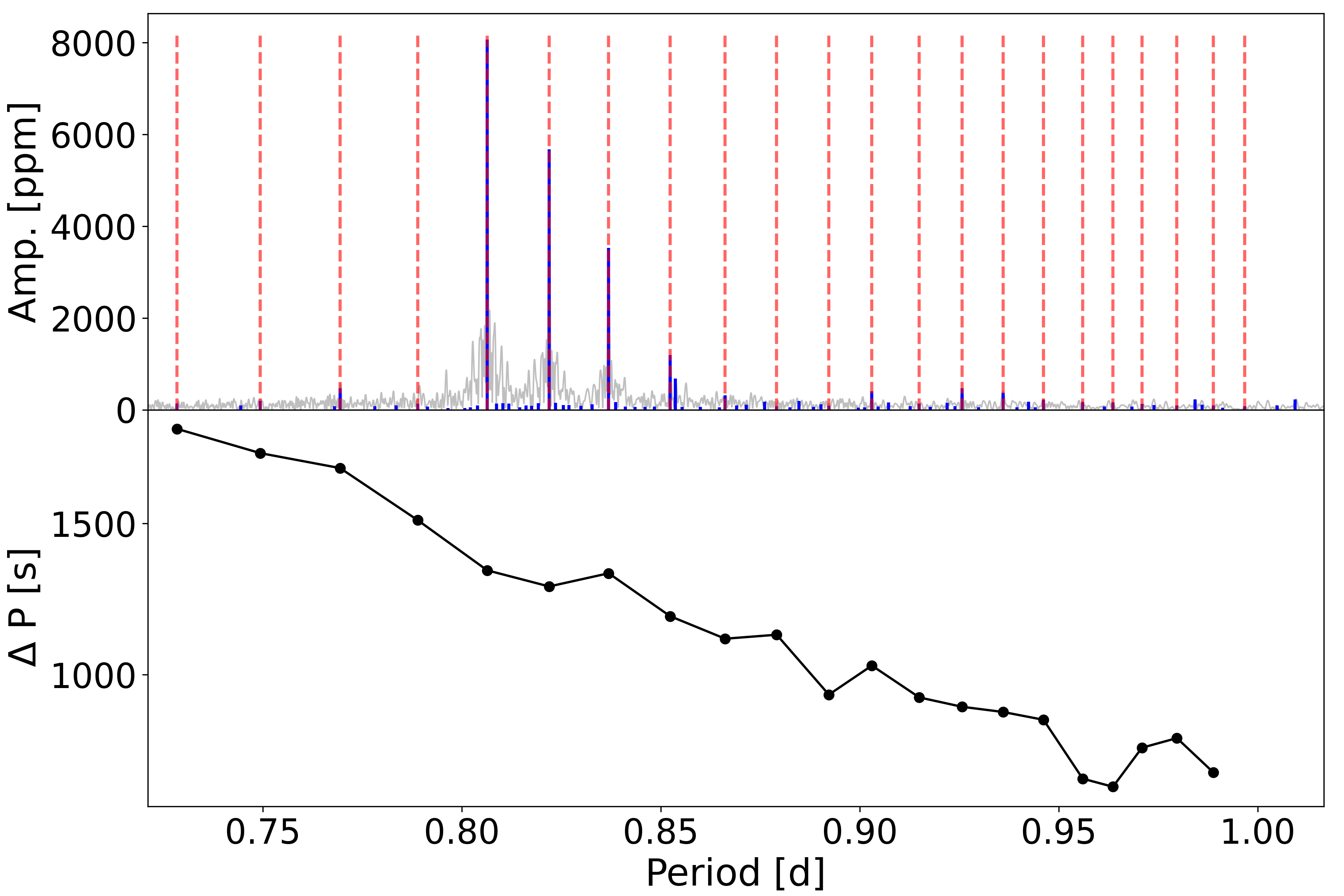}
		\caption{Prograde dipole g-mode pattern of KIC\,4930889. The top panel shows the amplitude spectrum in grey and the frequencies extracted by \citet{2021A&A...655A..59V} in blue. The frequencies selected to be part of the prograde dipole mode pattern are indicated by dashed red lines.
			The bottom panel shows the period-spacing pattern ($\Delta \rm P_n\equiv P_{n+1} - P_n$) of the selected prograde dipole modes.}
		\label{fig:KIC4930889_obs_pattern}
	\end{figure}

	The analysis of the \textit{Kepler} light curve and the subsequent frequency extraction was revisited by \citet{2021A&A...655A..59V}. We used the frequency list obtained by their `strategy 3' for the prewhitening procedure, as this method explains the highest fraction of variance for this light curve. This full list of frequencies is provided in \cref{tab:complete_freq_list} . The mode period pattern that we identified from this frequency list consists of 22 prograde dipole modes of consecutive radial order, which are listed in \cref{tab:freq_list} and shown in \cref{fig:KIC4930889_obs_pattern}. This provides us with the asteroseismic input for our modelling, $\Y{obs}$ composed of the individual mode frequencies $\text{Y}_i^{\text{Obs}}$ with $i=1, \ldots, 22$. Two additional modes occur in our pattern, while they were not present in the one based on the 20 modes found by \citet{2017A&A...598A..74P}. These are the two modes with the longest periods in \cref{fig:KIC4930889_obs_pattern}.

\begin{figure}[t!]
		\centering
		\includegraphics[width=\hsize]{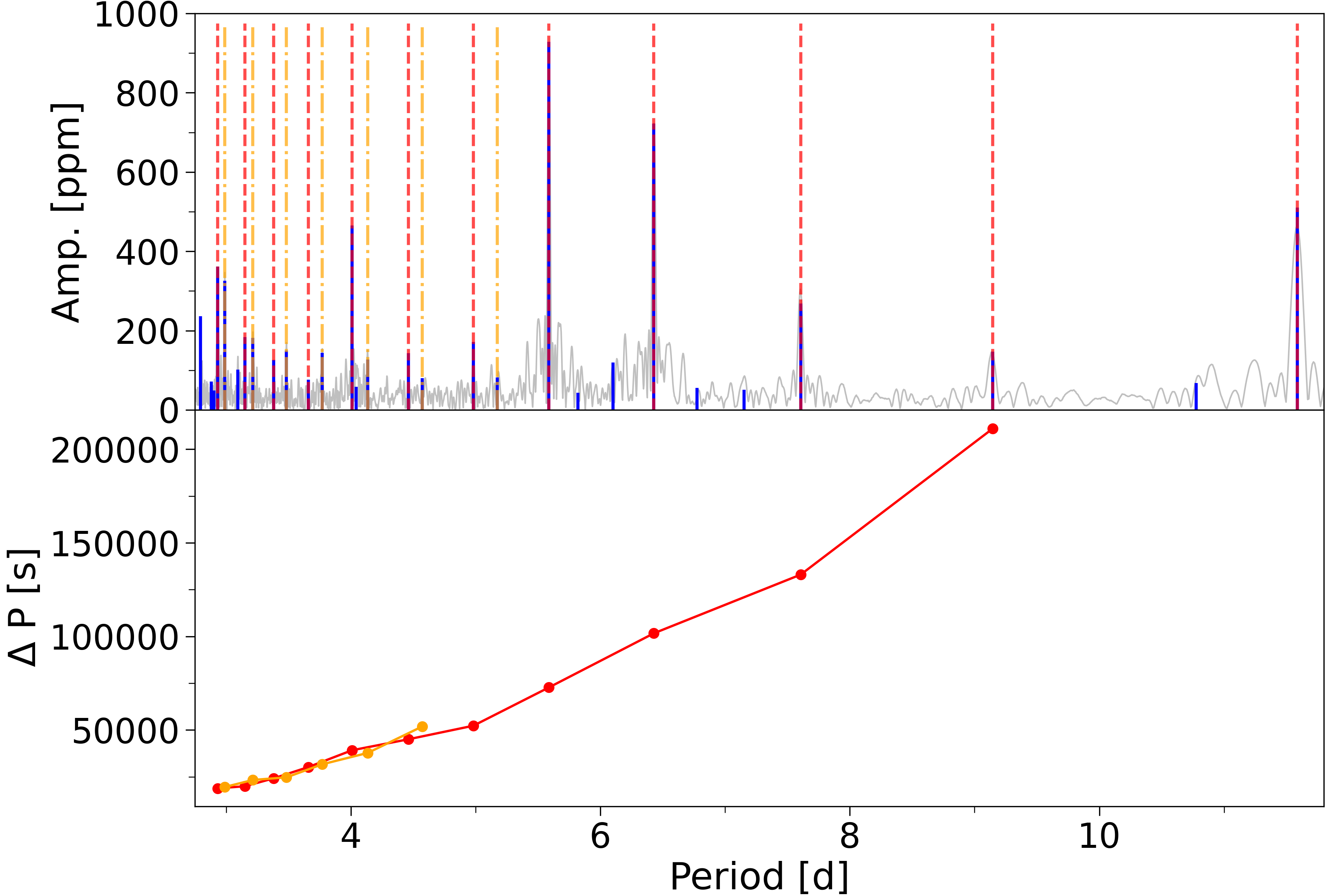}
		\caption{ Same as \cref{fig:KIC4930889_obs_pattern}, but for the possible additional series of the low frequency peaks.}
		\label{fig:KIC4930889_obs_pattern_additional}
	\end{figure}

	We additionally found a high amount of modes at lower frequencies that form two similar period series, listed in \cref{tab:freq_list_additional} and shown in \cref{fig:KIC4930889_obs_pattern_additional}. They have an upward tilt, typical for retrograde modes, and were also detected by \citet{2017A&A...598A..74P}. We also found one isolated peak in the amplitude spectrum at a period of 18.297$\pm$0.014$\,$d (f$_{165}$ in \cref{tab:complete_freq_list} with frequency 0.05465$\pm$0.00004$\,$d$^{-1}$), which is in perfect agreement with the orbital period found by \citet{2017A&A...598A..74P}. 
	We searched the full list of detected frequencies in \cref{tab:complete_freq_list} for frequencies that correspond to multiples of the orbital frequency within a 2$\sigma$ uncertainty interval. 
	Frequencies f$_{160}$ ($0.10934\pm0.00005$d$^{-1}$), f$_{155}$ ($0.16388\pm0.00006$d$^{-1}$), f$_{150}$ ($0.21865\pm0.00008$d$^{-1}$), and f$_{144}$ ($0.27325\pm0.00009$d$^{-1}$) coincide with two, three, four, and five times the orbital frequency, respectively. f$_{160}$ and f$_{144}$ correspond to f$_{11}$ and f$_4$ from the first additional pattern in \cref{tab:freq_list_additional}, and f$_{150}$ corresponds to f$_6$ from the second additional pattern \cref{tab:freq_list_additional}.
	Frequencies f$_{115}$ ($0.81955\pm0.00013$d$^{-1}$) and f$_{70}$ ($1.09295\pm0.00005$d$^{-1}$) coincide with 15 and 20 times the orbital frequency, respectively, with f$_{70}$ corresponding to f$_{13}$ from the main dipole mode period series in \cref{tab:freq_list}. The latter of these two may not be a tidally excited oscillation, since coinciding with a relatively high multiple of the orbital frequency could be coincidental, and since it falls in line with the dipole mode period series. 
	
	Since some of the signals in the secondary patterns are low multiples of the orbital frequency, those signals might be caused by proximity effects instead of actual pulsations, and those patterns as a whole are in any case likely influenced by the binary orbit.
	Additionally, we could not unambiguously determine the degree of the involved modes in these patterns from our single-star approach to the asteroseismology. We therefore do not include them in our modelling at this stage of the work. These two extra patterns offer potential for future more in-depth modelling based on close binary evolution models. Inclusion of these patterns requires developing a dedicated method to include tidal interactions, which is beyond the scope of the current study. Here, we restrict ourselves to asteroseismology based on single-star evolution models, as discussed in \cref{sec:computations}.
	
	\section{Orbital modelling of proximity effects}
	
\begin{figure}
		\includegraphics[width=\hsize]{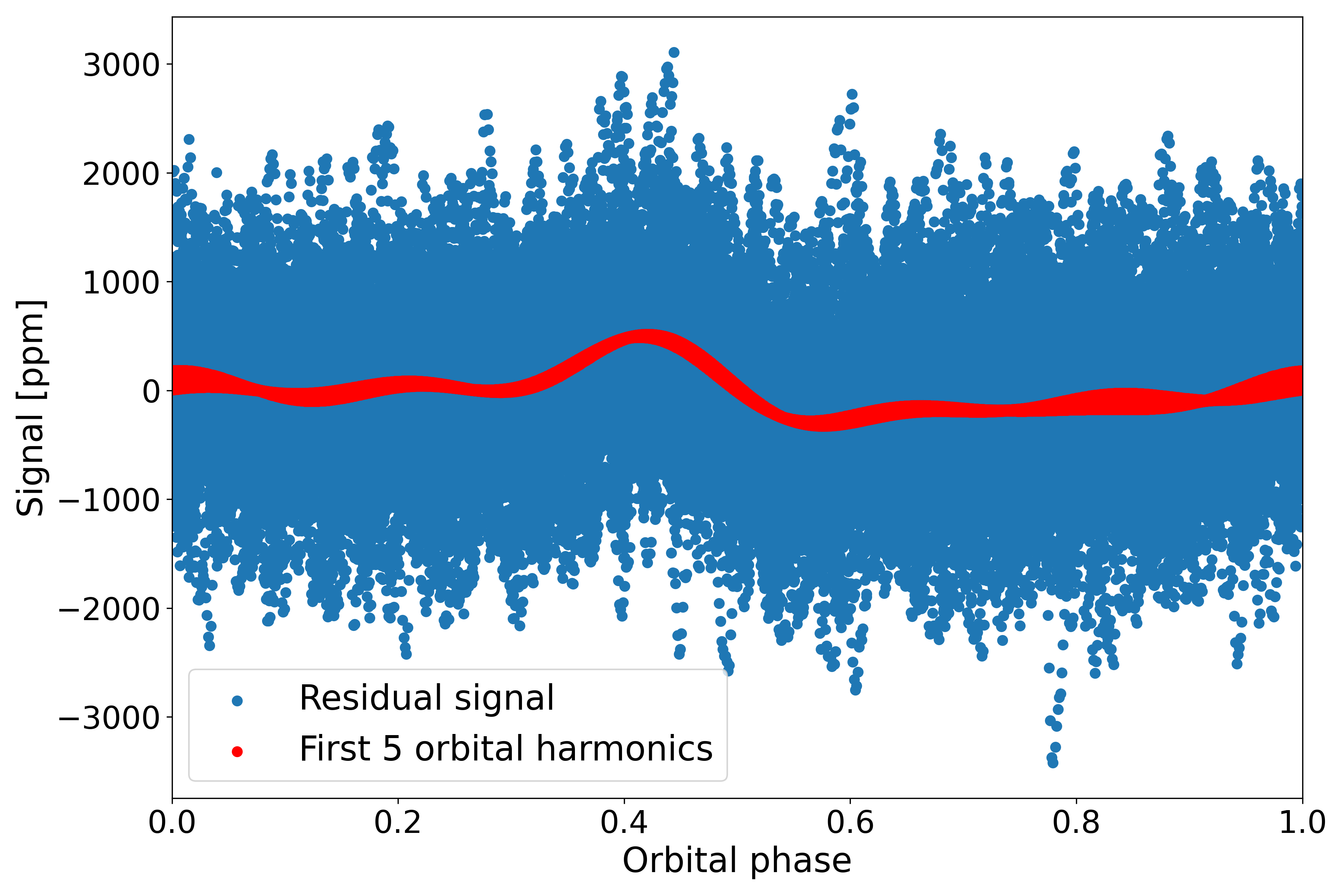}
		\caption{Phase-folded residual light curve. Prewhitened for all frequencies listed in \cref{tab:complete_freq_list}, except those corresponding to the first five orbital harmonics. The signal from those first five orbital harmonics is shown in red.}
		\label{fig:residuals}
	\end{figure}

	\Cref{fig:residuals} shows the phase-folded light curve after prewhitening for all detected significant frequencies apart from the multiples of the orbital frequency.
	The residual intrinsic variability in the light curve that is left after the stopping criterion employed by \citet{2021A&A...655A..59V} is dominant over the one caused by the orbital motion.
	We employ \texttt{PHOEBE} version 2.4.11 \citep{2021AAS...23714003C} to construct models for the orbital harmonics. We fix the mass ratio, eccentricity, argument of periapsis, and projected semi-major axis to the values listed in \cref{tab:spectro_parameters_pap}. Furthermore we fix the orbital period to 18.297\,d, since this is the orbital frequency retrieved from the light curve itself, it falls within the uncertainty of the value in \cref{tab:spectro_parameters_pap}, and the phase-folded light curve has less scatter than if 18.296\,d from \cref{tab:spectro_parameters_pap} is used in the phase fold. The effective temperatures are fixed to the values from \cref{tab:spectro_parameters_jon} since they only have a minor impact on the simulated light curve compared to the other parameters. The surface gravity of the stars are left as free parameters, with the values from \cref{tab:spectro_parameters_jon} as initial guesses. The inclination is a free parameter as well, as no prior information is available on its value. 

\begin{figure*}[tp]
		\centering 
		\begin{subfigure}{.33\textwidth}
			\includegraphics[width=\hsize]{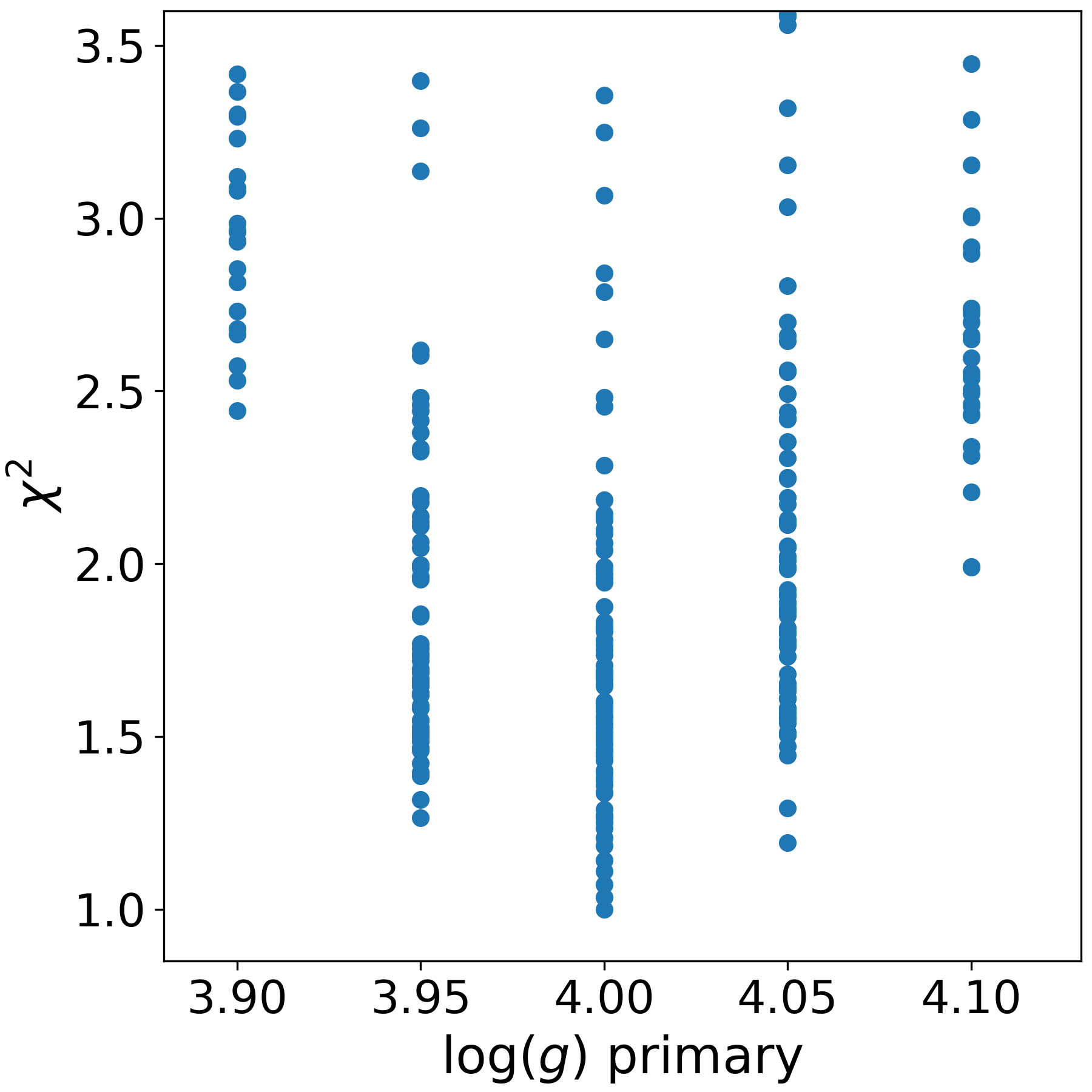}
		\end{subfigure}
		\begin{subfigure}{.33\textwidth}
			\includegraphics[width=\hsize]{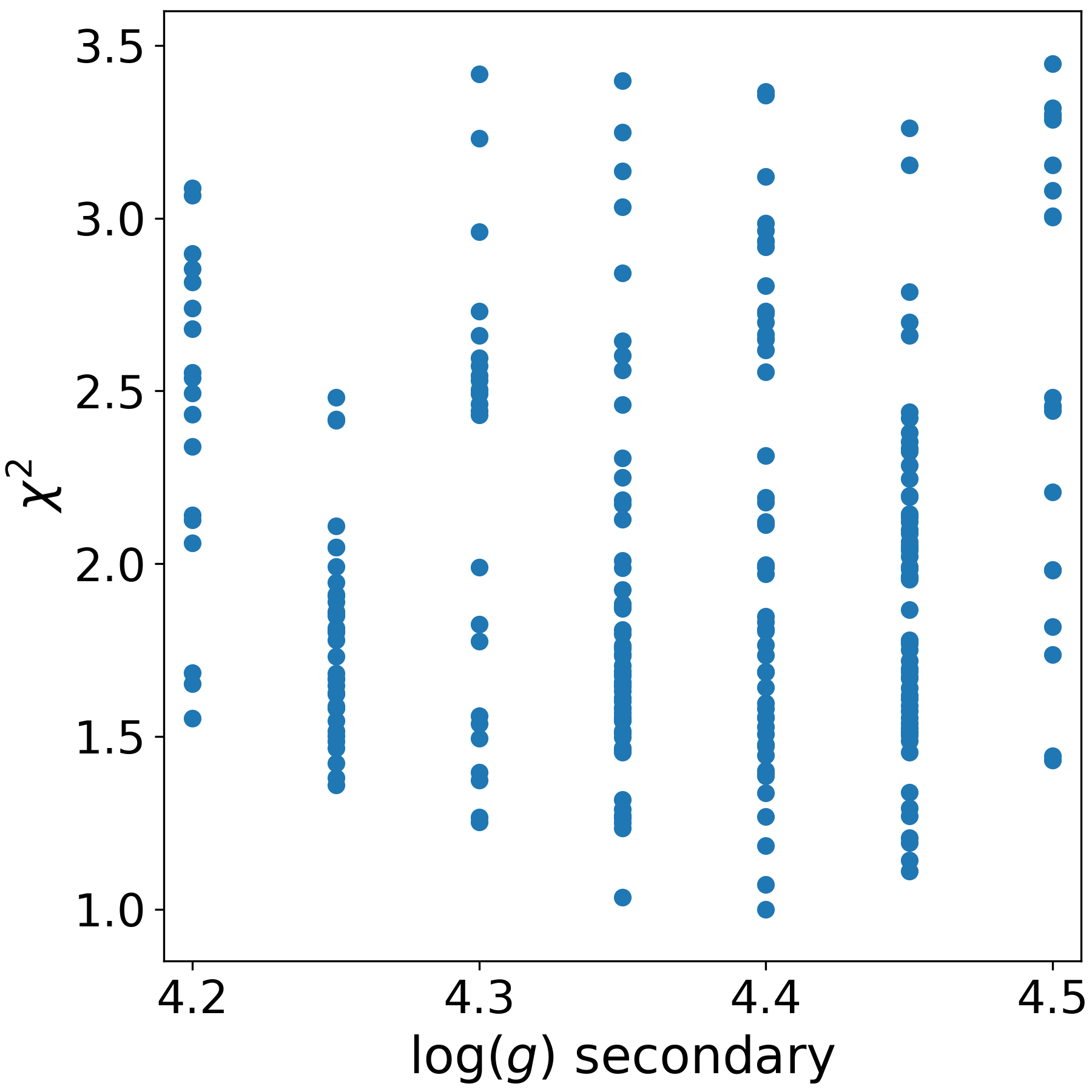}
		\end{subfigure}
		\begin{subfigure}{.33\textwidth}
			\includegraphics[width=\hsize]{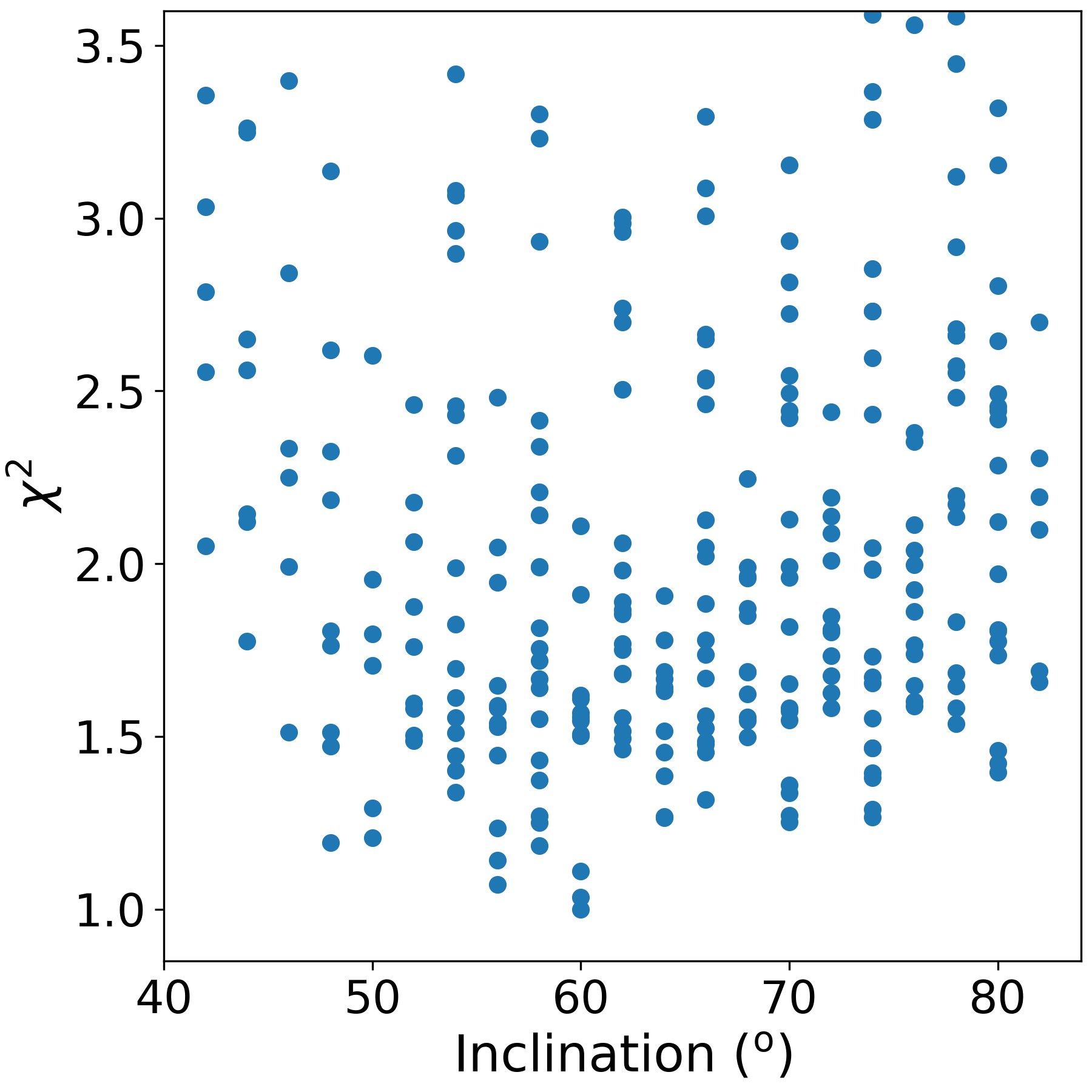}
		\end{subfigure}
		\caption{ $\chi^2$ values of the \texttt{PHOEBE} models. The values are projected along the log(g) of the primary (left), the log(g) of the secondary (center), and along the inclination of the binary (right). }
		\label{fig:chi2}
	\end{figure*}

\begin{figure*}
	\begin{minipage}[c]{0.70\textwidth}
		\includegraphics[width=\textwidth]{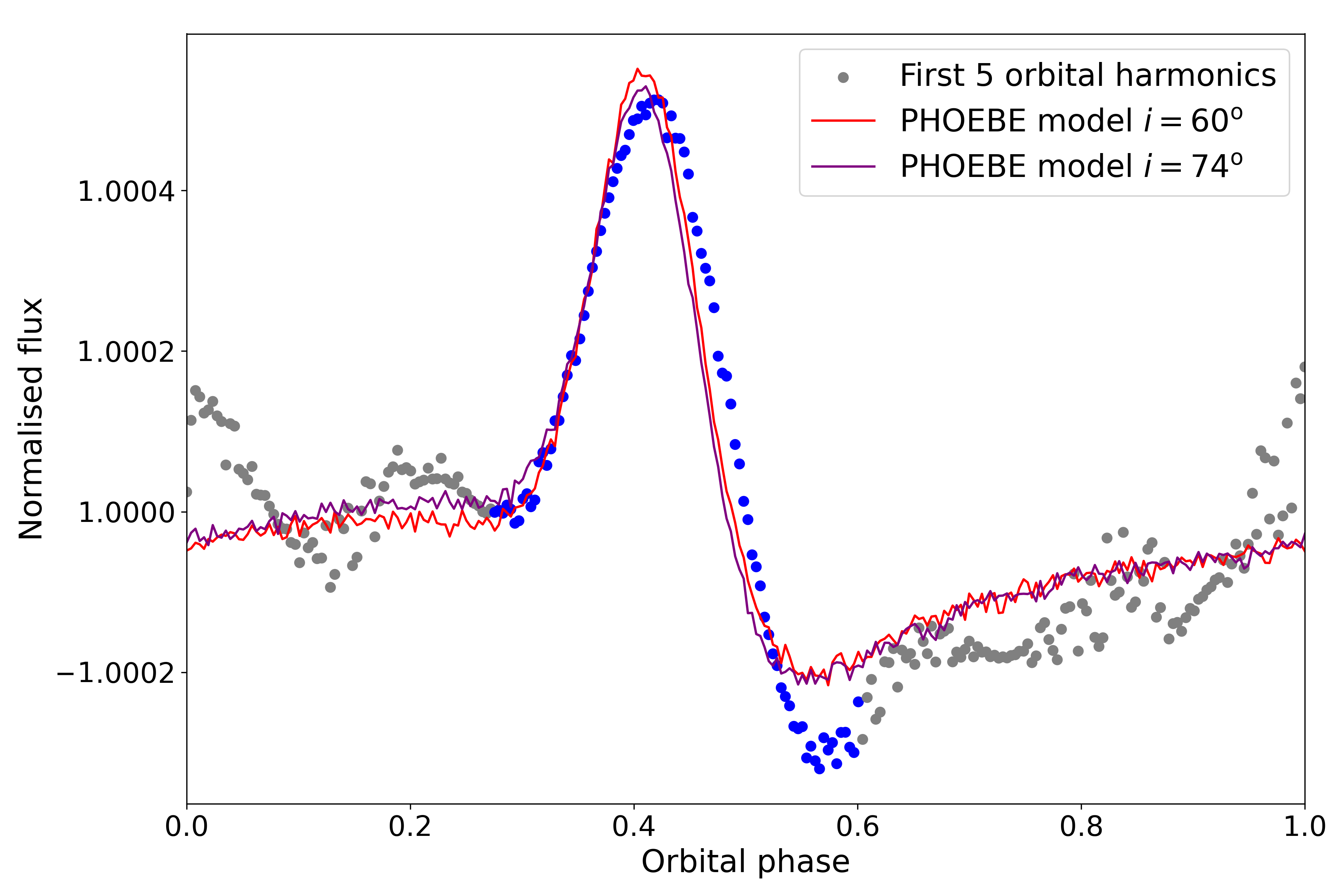}
	\end{minipage}\hfill
	\begin{minipage}[c]{0.30\textwidth}
		\caption{ Phase-folded \texttt{PHOEBE} model for the five orbital harmonics from \citet{2021A&A...655A..59V}. The dots in blue are included in the mask and modelled in the parameter study, whereas the dots in grey are excluded by the mask.}
 		\label{fig:Phoebe_model}
	\end{minipage}
\end{figure*}

	The Nelder-Mead algorithm \citep{10.1093/comjnl/7.4.308} is employed to optimise this initial setup. 
	Afterwards we compute a small parameter study around the retrieved solution, of which the projected goodness of fit can be seen in \cref{fig:chi2}. 
	As discussed before, the leftover intrinsic variability in the light curve is dominant, we therefore employ a mask during this parameter study to only model the bump in the light curve around orbital phase 0.4 since this is the clearest signal that is present in the harmonics.
	As can be seen from \cref{fig:chi2}, the best fitting surface gravity of the primary star deviates from the value of $3.55\pm0.24$ from \citet{2019MNRAS.482.1231J}. Their value of $4.38\pm0.10$ for the surface gravity of the secondary agrees very well with the best fitting values we retrieve in \cref{fig:chi2}. The distribution for the $\chi^2$ values of the inclination, as shown in \cref{fig:chi2}, is however much flatter than those of the surface gravities. \Cref{fig:Phoebe_model} demonstrates that inclination is indeed not very well constrained. The figure shows models with inclination angles of 60$^\circ$ and 74$^\circ$, which both reproduce the modelled signal comparably well. The absence of a clear minimum in the $\chi^2$ distribution of the models, combined with the wide range of inclinations that produce visually similarly good models, leaves us unable to confidently use this as a constraint on our asteroseismic modelling. 
		
	%%%%%%%%%%%%%%%%%%%%%%%%%%%%%%%%%%%%%%%%%%%%%%%%%%%%%%%%%%%%%%%%%%%%%%%%%%%%%%%%%%%%%%%%%%%%%%
	\section{Computation of theoretical mode frequencies} \label{sec:computations}
	%--------------------------------------------------------------------------------------------
	\subsection{Stellar equilibrium models}
\begin{figure*}
		\begin{minipage}[c]{0.69\textwidth}
			\includegraphics[width=\hsize]{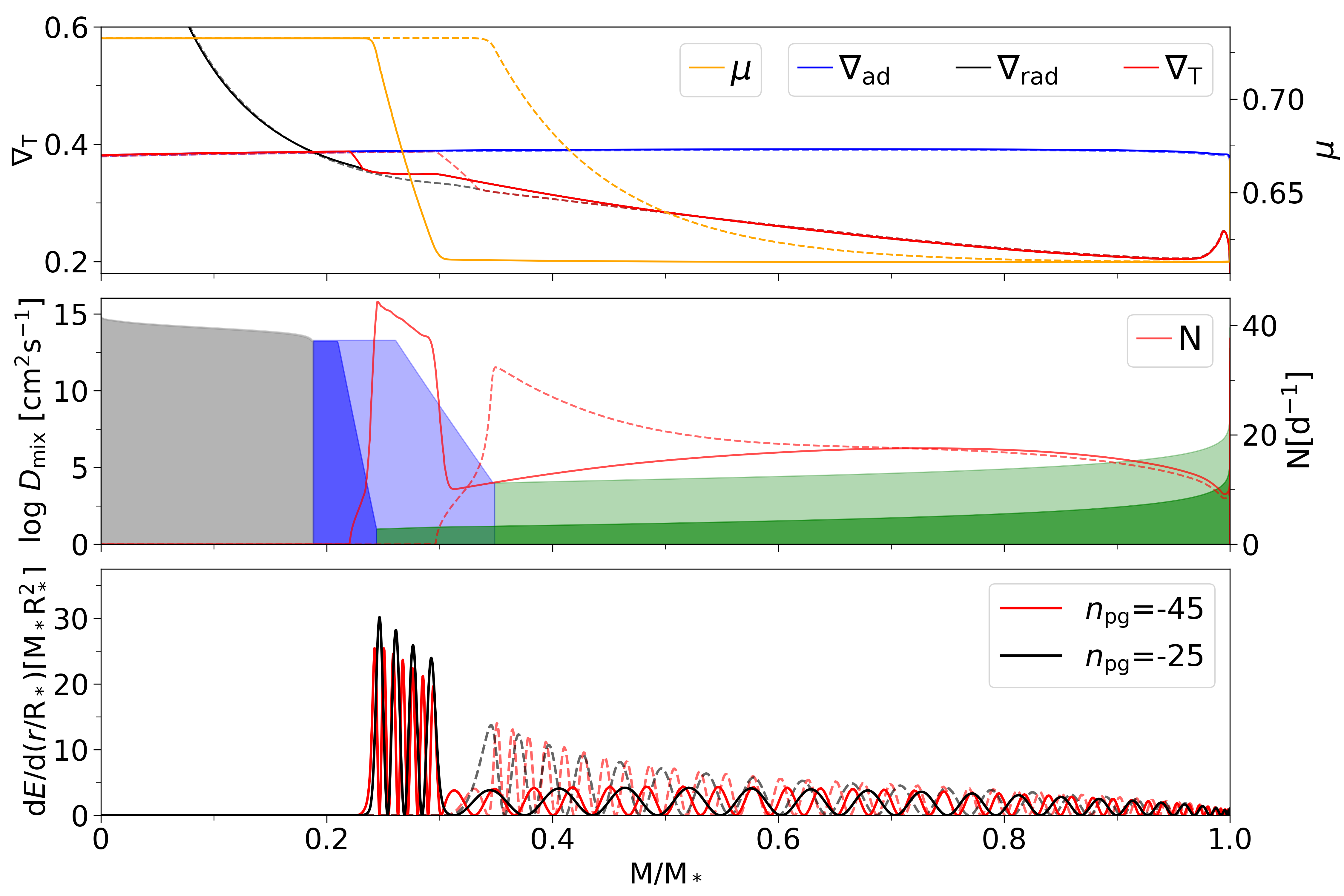}
		\end{minipage}\hfill
		\begin{minipage}[c]{0.3\textwidth}
		\caption{Radial profiles of a 4\msol \:star from the P\'eclet grid with a central hydrogen content $X_{\rm c}=0.5$. The top panel shows the temperature gradients and the mean molecular weight per gas particle ($\mu$). The middle panel shows the Brunt-V\"ais\"al\"a frequency (N), as well as the shape of the mixing profiles, divided in convective core (grey), near-core mixing (blue), and diffusive mixing in the outer radiative envelope (green). The bottom panel shows the mode inertia of two g modes with different radial orders. All dashed lines and transparent colours correspond to a model with the maximum amount of mixing included in our grid ($\acbm=0.3,\, \fcbm=0.03,\, \log(\D[env])=4$), whereas the solid lines and darker colours correspond to a model with a considerably lower amount of mixing  ($\acbm=0.1,\, \fcbm=0.01,\, \log(\D[env])=1$).}
			\label{fig:mixing_profile}
		\end{minipage}
	\end{figure*}

	Following a similar setup as \citet{2021A&A...650A.175M}, we compute two grids of single star models as input for the pulsation computations. The input physics for these two grids is the same, apart from the adopted temperature gradient in the core CBM region. The first, called the radiative grid, adopts the radiative temperature gradient in that transition zone. The other is termed the  P\'eclet grid and adopts a temperature gradient based on the P\'eclet number in that transition zone, following the same prescription as in Eq.\,(5) of \citet{2021A&A...650A.175M}. This prescription includes a convective penetration zone extending the convective core, which entails that (at least a part of) the CBM region is fully adiabatic. The mixing coefficient in the CBM region is governed by two parameters, $\acbm$ and $\fcbm$ which dictate the step-like and exponentially decaying parts of the region respectively. The diffusive mixing in the radiative envelope is implemented to increase going further outwards due to internal gravity waves as deduced by \citet{2017ApJ...848L...1R}, following \citet{2018A&A...614A.128P}. The level of this mixing in the radiative envelope at its inner boundary with the CBM region is set by the parameter $\D[env]$.
	The parameter ranges for the two grids with different temperature gradients are identical and listed in \cref{tab:parameters}. We note that the upper bound of the central hydrogen fraction is the initial fraction at the zero-age main sequence, which can vary depending on the initial metallicity of that model. \Cref{fig:mixing_profile} illustrates the stellar structure of a model with the maximum amount of mixing included in our grid, and compares it with the structure of one of the models with considerably less mixing. We can clearly see that strong mode trapping occurs in the CBM region when the amount of mixing is on the lower end. An increased amount of mixing in the envelope causes the chemical gradient to be less steep, entailing a much less pronounced peak of the Brunt-V\"ais\"al\"a frequency, and a greatly reduced (or even absent) mode trapping.

%	\begin{figure*}[ht]
%		\centering
%		\includegraphics[width=\hsize]{figures/profile_DO_Z0.012_M4.00_Xc0.50.png}
%		\caption{Radial profiles of a 4\msol \:star from the P\'eclet grid with a central hydrogen content $X_{\rm c}=0.5$. The top panel shows the temperature gradients and the mean molecular weight per gas particle ($\mu$). The middle panel shows the Brunt-V\"ais\"al\"a frequency (N), as well as the shape of the mixing profiles, divided in convective core (grey), near-core mixing (blue), and diffusive mixing in the outer radiative envelope (green). The bottom panel shows the mode inertia of two g modes with different radial orders. All dashed lines and transparent colours correspond to a model with the maximum amount of mixing included in our grid ($\acbm=0.3,\, \fcbm=0.03,\, \log(\D[env])=4$), whereas the solid lines and darker colours correspond to a model with a considerably lower amount of mixing  ($\acbm=0.1,\, \fcbm=0.01,\, \log(\D[env])=1$).}
%		\label{fig:mixing_profile}
%	\end{figure*}

	The two grids of stellar evolution models are computed using the stellar evolution code \texttt{MESA} \citep{2011ApJS..192....3P,2013ApJS..208....4P,2015ApJS..220...15P,2018ApJS..234...34P,2019ApJS..243...10P} version r15140.
	The models employ an Eddington grey atmosphere as atmospheric boundary condition and make use of the OP opacity tables \citep{2005MNRAS.362L...1S}. They contain the standard chemical mixture of OB stars in the solar neighbourhood deduced by \citet{2012A&A...539A.143N} and \citet{2013EAS....63...13P}.
	We determine the initial helium fraction by adopting an enrichment law $Y_{\rm ini} = Y_{p} + (\Delta Y/\Delta Z) Z_{\rm ini}$. We set the primordial helium abundance $Y_p = 0.2465$, as determined by \citet{2013JCAP...11..017A}. Since there is currently no consensus on the value of $\frac{\Delta Y}{\Delta Z}$ \citep[e.g.][and references therein]{2019MNRAS.483.4678V}, we require that the galactic enrichment ratio, $\frac{\Delta Y}{\Delta Z}$, is able to reproduce the mass fractions of the adopted chemical mixture ($X$=0.71, $Y$=0.276, $Z$=0.014) derived by \citet{2012A&A...539A.143N}.
	This leads us to adopt $\Delta Y/\Delta Z$=2.1. After $Y_{\rm ini}$ is determined according to this enrichment law, $X_{\rm ini}$ is set following $X_{\rm ini}=1-Y_{\rm ini}-Z_{\rm ini}$.

	We adopt the mixing length theory as developed by \citet{1968pss..book.....C} with a mixing length parameter $\alpha_{\text{mlt}}=2.0$, and use the Ledoux criterion for convection without allowing for semi-convection. This is warranted since this form of slow mixing is absent in the presence of CBM \citep[e.g.][]{2020MNRAS.496.1967K}, which is included in the vast majority of our models. The exact location where the transition from core to near-core mixing is made, is determined by the $f_0$ parameter in \texttt{MESA}. We fix $f_0=0.005$, except for setting $f_0=0$ in the models where both $\acbm$ and $\fcbm$ are equal to zero as there is no CBM region for this case.
	A link to the detailed \texttt{MESA} setup is provided in \cref{appendix:inlist}.

	\begin{table}
		\caption{Parameter ranges of each of the two grids of equilibrium models used for the asteroseismic modelling, containing a total of 1191680 models per grid.}
		\label{tab:parameters}
		\centering
		\begin{tabular}{l l l l}
			\hline
			\hline
			Parameter & lower boundary & upper boundary & step size \\
			\hline
			$M_{\rm ini}$ [\msol] & 3.0 & 4.5 & 0.1 \\
			$Z_{\rm ini}$ & 0.008 &  0.024 & 0.004 \\
			$\acbm$  & 0 & 0.3 &  0.05 \\
			$\fcbm$   & 0 & 0.03 & 0.005 \\
			log($\D[env]$) & 0 & 4 &  1 \\
			$X_{\rm c}$ & 0.1 & $X_{ini}$ &  0.01 \\
			\hline
		\end{tabular}
	\end{table}

	%--------------------------------------------------------------------------------------------
	\subsection{Pulsation computations}

	The pulsation mode properties of the \texttt{MESA} equilibrium models are computed employing the stellar oscillation code \texttt{GYRE} \citep{2013MNRAS.435.3406T,2018MNRAS.475..879T}, version 6.0.1.
	Since non-adiabatic effects mainly become important in the outer stellar envelope, the adiabatic approximations are sufficient for our modelling work due to the mode inertias of the g modes being dominant near the stellar core. For computational reasons and given that it does not affect the mode frequencies at the level of measurement errors, we only perform non-adiabatic computations for some of our best models after the forward modelling is finished, in order to evaluate their mode excitation.
	We compute the dipole g modes for all our equilibrium models for an initial guess of the rotation frequency, assuming rigid rotation and relying on the traditional approximation of rotation \citep[TAR; e.g.][]{Eckart1960,Bildsten1996,Lee1997}, following its implementation in \texttt{GYRE} \citep[as described in][sec. 4]{2020MNRAS.497.2670T}.

	The stellar rotation frequency required to optimally reproduce the observed stellar pulsations differs for the varying equilibrium models. We therefore start from the same initial guess for each equilibrium model and rescale the g-mode frequencies for each model separately, following the TAR and assuming rigid rotation, to reproduce the observed pulsations as closely as possible. This optimisation is performed using the Levenberg-Marquardt method implemented in \texttt{LMFIT} \citep{2020zndo...3814709N}.
	To reduce the chances of the optimisation method returning a local minimum, we start the optimisation procedure from two separate initial values for the rotation. The first one being the initial guess, $\omega_{\text{initial, 1}} =\omega_{\text{guess}}$, used to calculate the \texttt{GYRE} model. The second initial value is taken by adjusting the first one by twice the difference between the initial value and its solution, so $\omega_{\text{initial, 2}} = \omega_{\text{guess}} - 2\cdot (\omega_{\text{guess}} - \omega_{\text{optimised, 1}}) = 2\cdot \omega_{\text{optimised, 1}} - \omega_{\text{guess}}$. This way the global minimum of the initial value problem is approached both from a higher and a lower initial value. In the case where these solutions do not converge, we take the best of the two returned solutions since it indicates that the other one returns a local minimum.

%	\begin{figure*}[ht]
%		\centering
%		\includegraphics[width=\hsize]{figures/rotation_rescaling.png}
%		\caption{ Rescaling period spacing patterns to an optimised rotation rate. Top panel shows the period-spacing pattern as calculated by \texttt{GYRE} with the initial guess for the rotation in blue, and the pattern rescaled to the optimised rotation frequency in orange. The inset figures are zoomed in on the region with the observed pulsations, with the modes selected to match the observations circled in red. The bottom panel shows the relative difference between the mode periods calculated by \texttt{GYRE} given the optimised rotation rate, and the mode periods from the rescaled pattern. The grey region denotes the observational uncertainties.}
%		\label{fig:rotation_rescaling}
%	\end{figure*}

	\begin{figure*}[ht]
	\begin{minipage}[c]{0.72\textwidth}
		\includegraphics[width=\hsize]{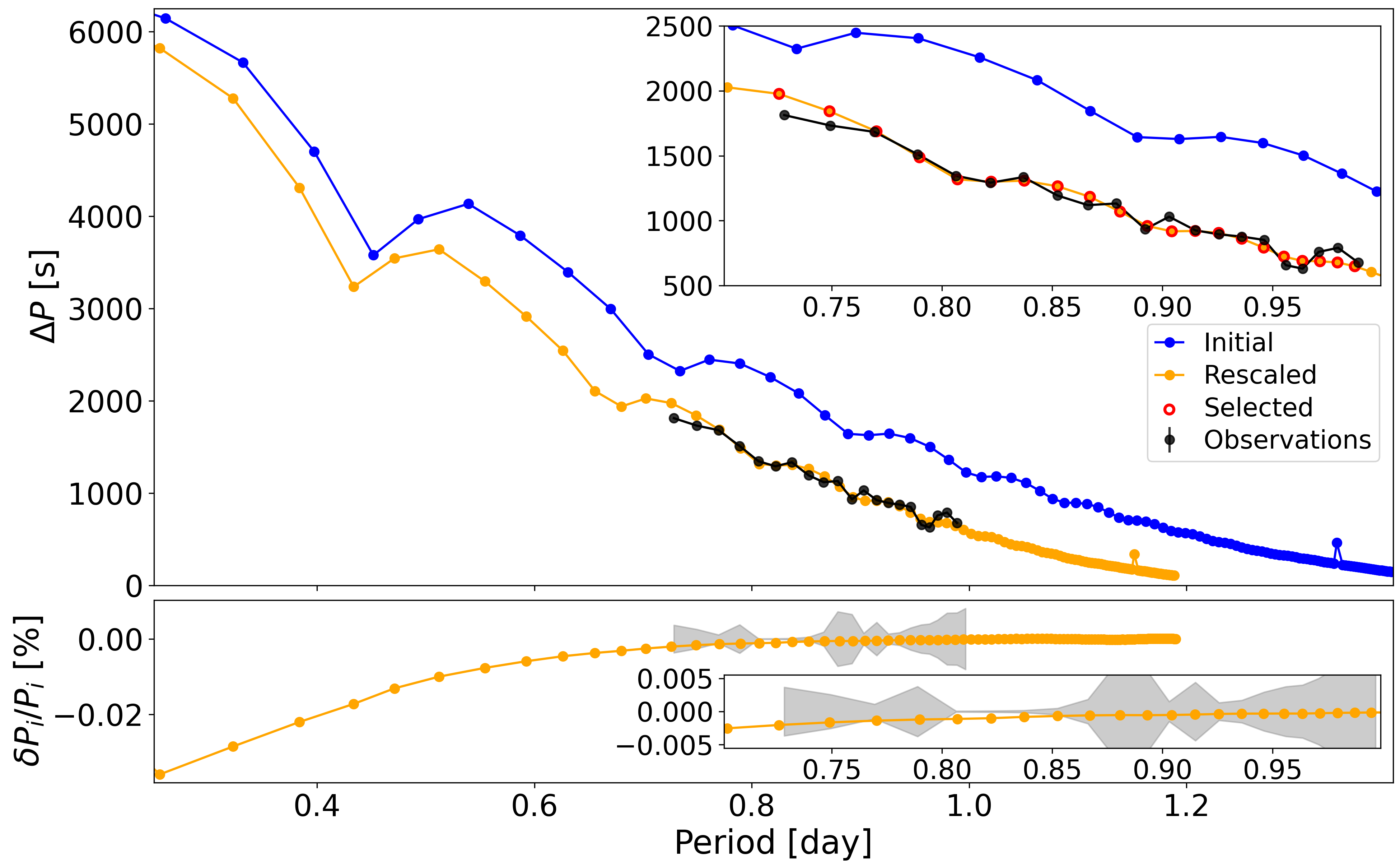}
	\end{minipage}\hfill
	\begin{minipage}[c]{0.275\textwidth}
		\caption{ Rescaling period spacing patterns to an optimised rotation rate. Top panel shows the period-spacing pattern as calculated by \texttt{GYRE} with the initial guess for the rotation in blue, and the pattern rescaled to the optimised rotation frequency in orange. The inset figures are zoomed in on the region with the observed pulsations, with the modes selected to match the observations circled in red. The bottom panel shows the relative difference between the mode periods calculated by \texttt{GYRE} given the optimised rotation rate, and the mode periods from the rescaled pattern. The grey region denotes the observational uncertainties.}
		\label{fig:rotation_rescaling}
	\end{minipage}
\end{figure*}

	\Cref{fig:rotation_rescaling} illustrates the rescaling of the period-spacing pattern due to a change in the rotation rate. It also shows the relative differences between the periods of the rescaled modes and the periods obtained by repeating the \texttt{GYRE} computation using that same optimised rotation rate.
	We find that the rescaled mode periods agree well with the periods computed by \texttt{GYRE} for the new rotation rate. The differences are of order 10$^{-3}\%$ in the asymptotic mode frequency regime where the observed pulsations occur, and even the largest differences at low radial orders are still relatively small (<0.05\%).
	Rescaling the g-mode frequencies to the optimised rotation frequency and selecting a set of the theoretical frequencies to match the observations yields for each equilibrium model a list of theoretically predicted dipole mode frequencies, $\Y{Theo}$ composed of $\text{Y}_i^{\text{Theo}}$, where $i$ stands for the radial order.
	The \texttt{GYRE} inlist to compute the initial frequency lists is provided through the link in \cref{appendix:inlist}.

	%%%%%%%%%%%%%%%%%%%%%%%%%%%%%%%%%%%%%%%%%%%%%%%%%%%%%%%%%%%%%%%%%%%%%%%%%%%%%%%%%%%%%%%%%%%%%%
	\section{Modelling approach}

	We utilise the same asteroseismic modelling procedure as \citet{2021A&A...650A.175M}. A brief overview is provided here for convenience without going too much into the details.

	\subsection{General mathematical framework} \label{sec:MD_Framework}

	We employ the Mahalanobis distance as a merit function for the maximum likelihood estimation in the asteroseismic modelling \citep[see][for its application to asteroseismic modelling]{2018ApJS..237...15A},

	\begin{align}
	\text{MD}_j = \lb \Y{theo}_{j} - \Y{obs} \rb ^T \lb V+\Sigma \rb ^{-1} \lb \Y{theo}_{j} - \Y{obs} \rb,
	\end{align}
	with $\Y{obs}$ the vector of observations and $\Y{theo}_{j}$ the corresponding vector of predicted values in gridpoint $j$. $\Sigma$ is the variance matrix due to the measurement errors of $\Y{obs}$ and $V$ is the variance--covariance matrix of $\Y{theo}$ capturing the theoretical uncertainties in the mode frequency predictions caused by the limited knowledge of the physical ingredients in the input physics of the equilibrium models, taking as well the correlations among the free parameters used to describe these ingredients into account.

	The modelling involves both statistical models that are non-nested, comparing models within one grid of equilibrium models, and statistical models that are nested, comparing equilibrium models across different grids where none, one, or both of the CBM parameters are fixed at zero. This allows for a comparison between different numbers of free parameters, including an evaluation whether the increase in goodness of fit outweighs the entailed punishment by the selection criterion for having an increased number of free parameters.

	We use the Akaike Information Criterion corrected for small sample size \citep[AICc,][Chapter\,2]{Claeskens2008} since it rewards fit quality but penalises complexity. It is defined as
	\begin{equation}\label{eq:aicc}
	\text{AICc} = -2\ln{\cal L} + \frac{2kN}{N-k-1},
	\end{equation}
	with $N$ and $k$ the number of observables and free parameters, respectively, and ${\cal L}$ the likelihood of a stellar model.
	In our framework for this star, $N=22$ when fitting periods, or 21 when fitting period spacings (that is the differences in period between two pulsation modes of consecutive radial order, $\Delta \rm P_n\equiv P_{n+1} - P_n$). The number of free parameters $k$ is 4, 5, or 6 depending on whether two, one, or zero of the CBM parameters ($\acbm, \fcbm$) are fixed in the nested grids. In case $k$=6, the list of parameters consists of ($M_{\rm ini}, Z_{\rm ini}, \acbm, \fcbm, \log(\D[env]), X_{\rm c}$).
	Rewriting the AICc for the likelihood function of the Mahalanobis Distance yields
	\begin{equation}
	\text{AICc}  = \ln(|V+\Sigma|) + k\ln(2\pi) + \text{MD}  + \frac{2kN}{N-k-1}.
	\end{equation}
	The performance of two nested models can be compared through their difference in AICc values $\Delta \text{AICc} = \text{AICc}_\text{A} - \text{AICc}_\text{B}$. Model B is preferred over model A if $\Delta \text{AICc} > 2$, with a (very) strong preference if $\Delta \text{AICc} > 6$ (10).

	We determine the uncertainty region of the best solution by employing Bayes' theorem, stating that the probability of a parameter $\theta^m$ occurring in the interval |$\theta^m_a$, $\theta^m_b$| is given by
	\begin{align}\label{eq:error_ellips}
	P(\theta^m_a < \theta^m < \theta^m_b | \mathbf{D}) =& \frac{\sum_i^q P(\mathbf{D}|\boldsymbol{\theta}_i) P(\boldsymbol{\theta}_i) }{\sum_j^Q P(\mathbf{D}|\boldsymbol{\theta}_j) P(\boldsymbol{\theta}_j) } \nonumber \\
	=& \frac{\sum_i^q P(\mathbf{D}|\boldsymbol{\theta}_i) \prod_l^k P({\theta}_i^l) }{\sum_j^Q P(\mathbf{D}|\boldsymbol{\theta}_j) \prod_l^k P({\theta}_j^l) }.
	\end{align}
	Index $j$ is summed over all $Q$ equilibrium models in the grid that are consistent within 3$\sigma$ of the spectroscopic $\log\,g$, $\Teff$, and stellar luminosity, that are also consistent with the observed metallicity and constraints from the binarity of the system.
	These binary constraints are explained in more detail in \cref{Isochrone-clouds} Index $i$ is summed over the $q$ models with the highest likelihood so that $P(\theta^m_a < \theta^m < \theta^m_b | \mathbf{D}) = 0.95$.

	We consider three approaches to match theoretical mode periods to the observed ones, and analyse the results of the method that performs best for each grid.
	In the first two we begin matching mode periods starting from the theoretical period that is closest to the either the mode with the highest observed amplitude or the highest-frequency detected in the observed pattern.
	The third option is to match each observed mode period to its best matching theoretical counterpart, and adopt the longest sequence of consecutive modes that we get in this way. The rest of the pattern is then build consecutively in radial order starting from this sequence. These three options of pattern construction will henceforth be referred to as highest amplitude, highest frequency, and longest sequence.

	Apart from just the mode periods, we also consider the period-spacing values as a set of observables to be used in our modelling procedure. The condition numbers of the variance-covariance matrices $V+\Sigma$ are used to determine the best of these sets of observables. The condition number $\kappa$ is defined as the ratio of its maximum to minimum eigenvalue,
	\begin{equation} \label{eq:cond_nr}
	\kappa(V+\Sigma) = \frac{|\lambda_{max}(V+\Sigma)|}{|\lambda_{min}(V+\Sigma)|}.
	\end{equation}
	This gives an indication of how well- or ill-conditioned the matrix is with respect to the inversion to be computed, with lower values being better conditioned.

	\subsection{Isochrone clouds\label{Isochrone-clouds}}

	The methodology from \citet{2021A&A...650A.175M} as summarised in Sect.\,\ref{sec:MD_Framework} considers the system's asteroseismic and spectroscopic data from a single-star perspective.
	KIC\,4930889 is however a double-lined spectroscopic eccentric $(e=0.32)$ binary. Hence, we can utilise the information obtained from the binarity of the system to put additional constraints on the models in an attempt to lift some of the degeneracies that are present. We employ the use of isochrone clouds \citep{2019MNRAS.482.1231J}, which is in this application the collection of isochrones of a given age but for all combinations of $\acbm$, $\fcbm$, $\log\D[env]$ present in our grid. Constructing an isochrone cloud coupled to a model of a certain grid, we enforce all models in that cloud to be of an age that differs less than one gridstep in age from this model, have the same initial metallicity, and have a mass that is compatible within the error margin of the mass ratio of the system (listed in \cref{tab:spectro_parameters_pap}).
	We computed some additional evolutionary tracks for masses above and below the grid range listed in \cref{tab:parameters} to allow all masses that we could expect for the companion from the observed mass ratios to be present in the isochrone clouds.

\begin{figure}[ht!]
		\centering
		\includegraphics[width=\hsize]{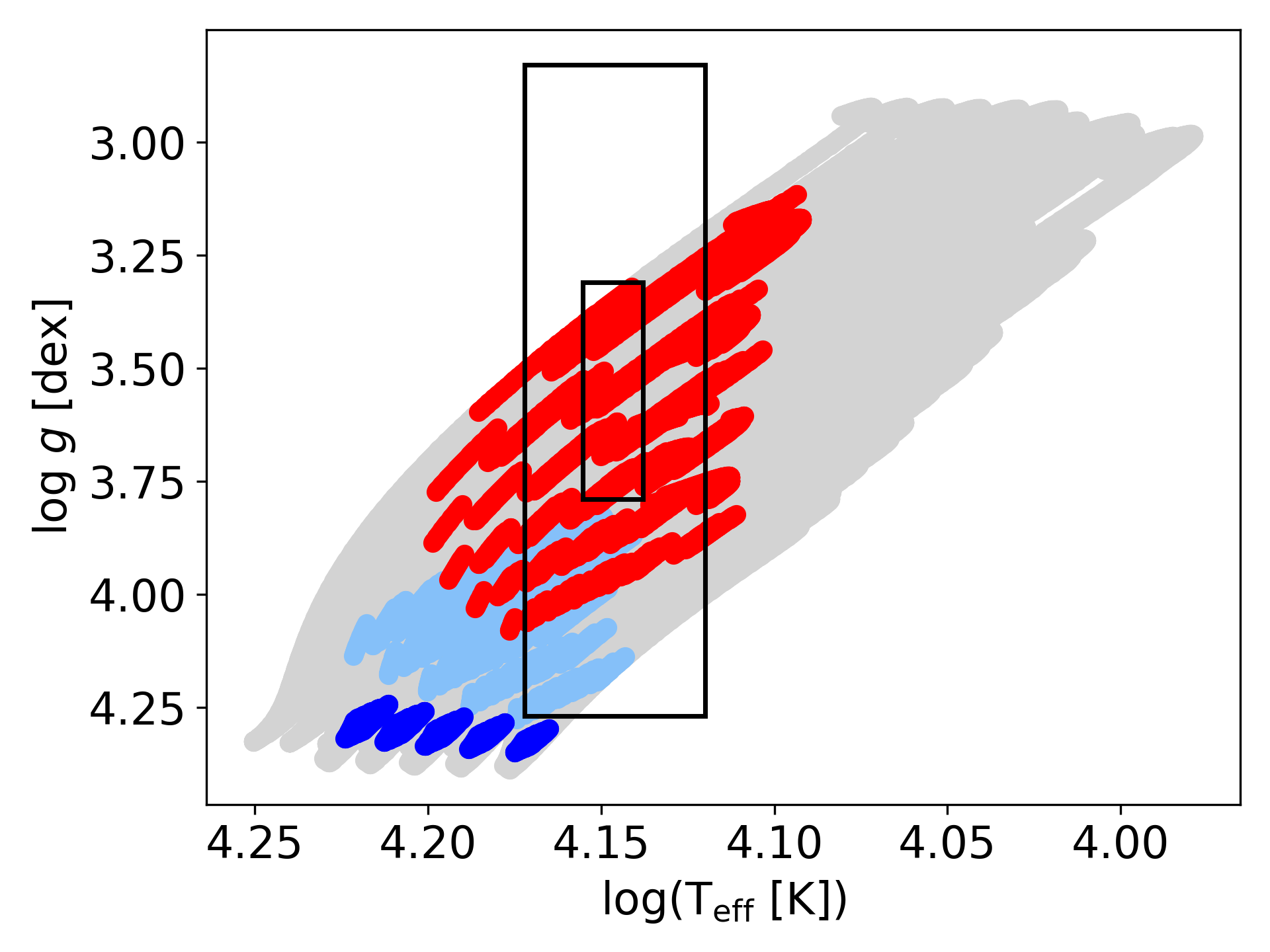}
		\caption{Isochrone clouds of the primary star matching three different models of the secondary. The black lines show the 1$\sigma$ and 3$\sigma$ spectroscopic $\Teff$ and $\log\,g$ error boxes of the primary star.}
		\label{fig:isocloud_secondary_concept}

	\end{figure}

	\Cref{fig:isocloud_secondary_concept} illustrates how the constraints of the isochrone clouds are applied for the case where we assume the secondary star to be the pulsator. It shows the isochrone clouds of the primary star for three different models of the secondary, which are arbitrarily chosen for the purpose of this visual representation.
	These three models are among the best, that is having the lowest AICc values, and they would have been included in the error ellipses if the system were modelled from a single-star perspective without any constraints from binarity.
	The light-grey tracks in the background show all models with the same metallicity and a mass within the observed mass ratio, and only the models that have an age difference smaller than one gridstep are shown in colour. Two of the isochrone clouds fall partially within the 1$\sigma$ or 3$\sigma$ errors of the companion star and are thus accepted as viable solutions. The third isochrone cloud falls completely outside of the 3$\sigma$ spectroscopic error region, and is hence not accepted as a solution. Although \cref{fig:isocloud_secondary_concept} only showcases the constraints on $\log g$ and $\Teff$, the stellar luminosity is also used in these isochrone-cloud constraints.

	\section{Modelling results}
	\begin{figure*}[ht!]
		\centering
		\includegraphics[width=0.7\hsize]{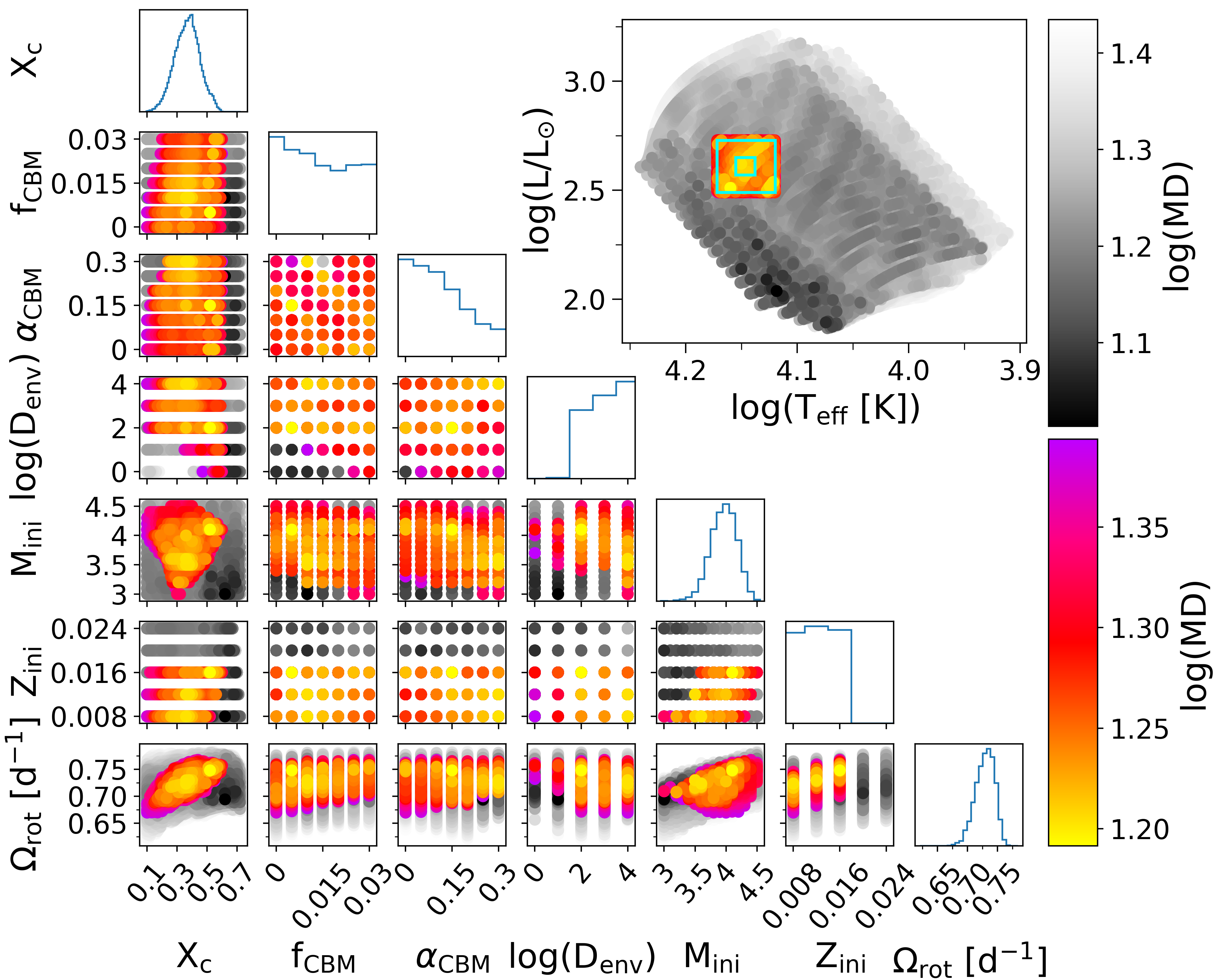}
		\caption{Corner plot for the radiative grid. Made using period spacings in a Mahalanobis distance merit function and spectroscopic and luminosity constraints from the primary star. The 50\% best models are shown, colour-coded according to the log of their merit function value (at right). The models in colour fall within the 2\,$\sigma$ error ellipse of the MD constructed using \cref{eq:error_ellips}, whilst the models in grey fall outside of this error ellipse. The figures on the diagonal show binned parameter distributions of the models in the error ellipse, and the panel at the top right shows a Hertzsprung–Russell diagram with the 1 and 3$\sigma$ $\Teff$ and $\log\,$L error boxes.}
		\label{fig:corr_KIC4930889A_full_DO_highest_frequency_MD_dP}
	\end{figure*}
	\begin{figure*}[ht!]
		\centering
		\includegraphics[width=0.7\hsize]{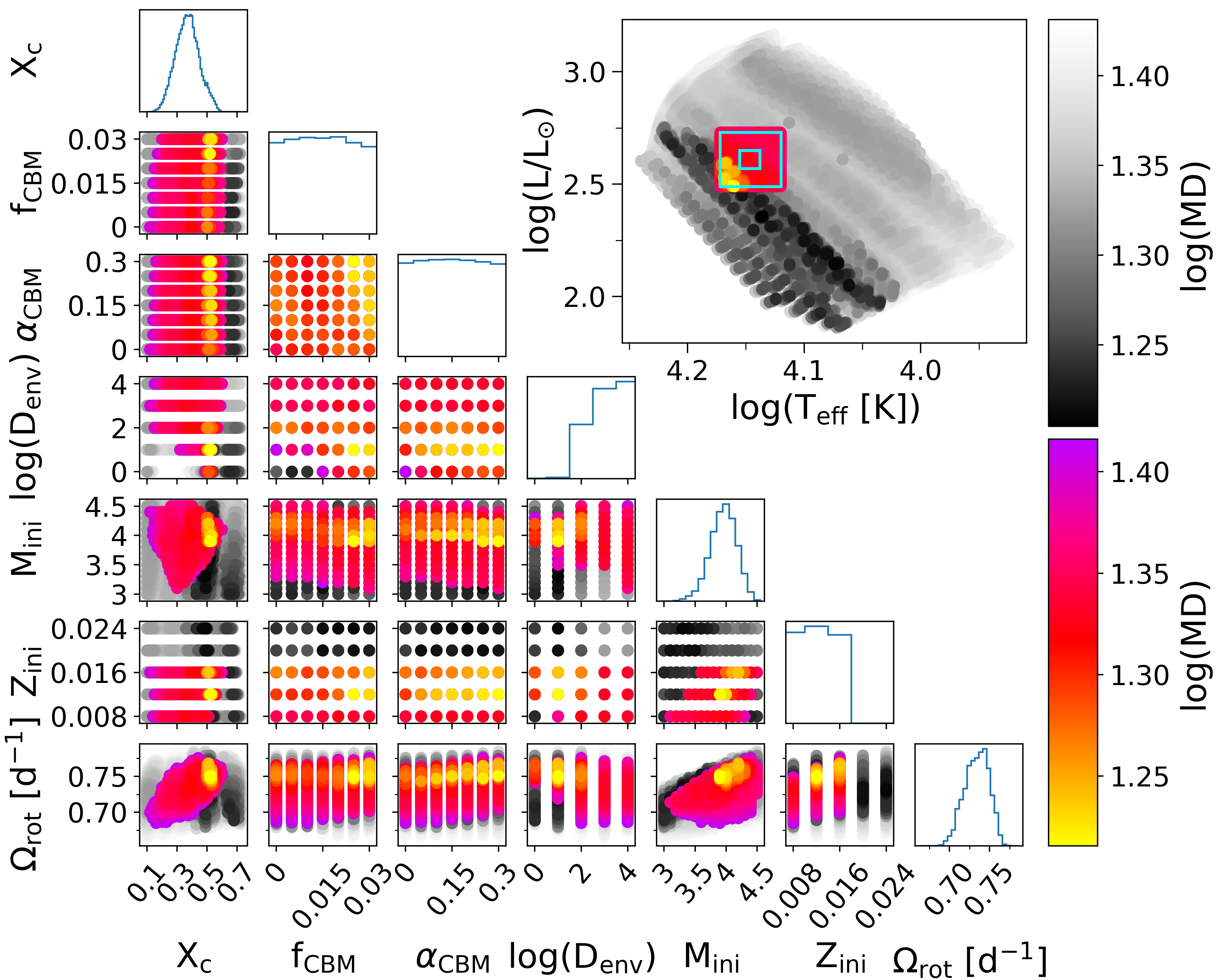}
		\caption{Corner plot (as in \cref{fig:corr_KIC4930889A_full_DO_highest_frequency_MD_dP}) for the P\'eclet grid. Made using period spacings in a Mahalanobis distance merit function and spectroscopic and luminosity constraints from the primary star.}
		\label{fig:corr_KIC4930889A_full_ECP_highest_frequency_MD_dP}
	\end{figure*}

	\begin{figure*}[ht!]
		\centering
		\includegraphics[width=0.7\hsize]{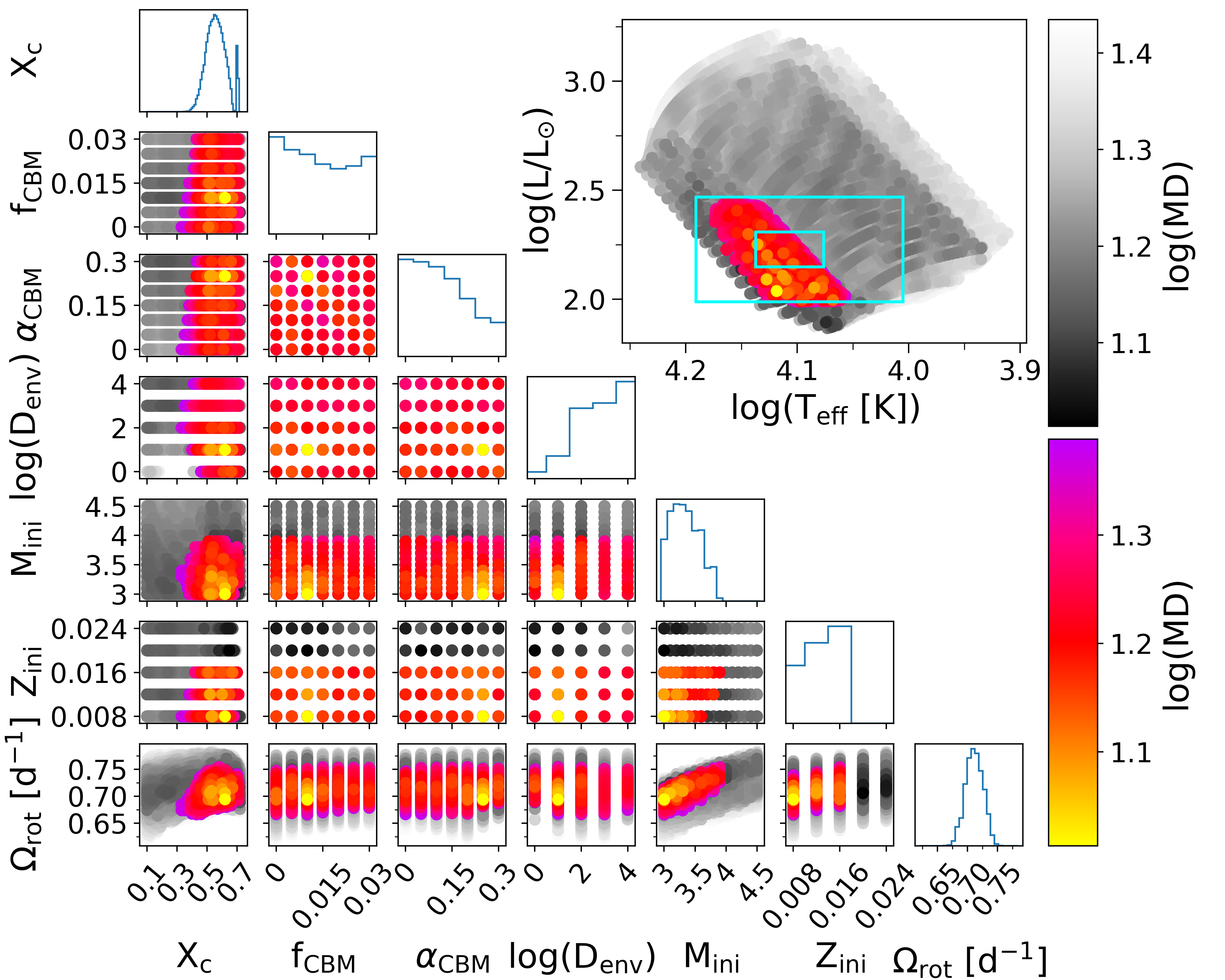}
		\caption{Corner plot (as in \cref{fig:corr_KIC4930889A_full_DO_highest_frequency_MD_dP}) for the radiative grid. Made using period spacings in a Mahalanobis distance merit function and spectroscopic and luminosity constraints from the secondary star.}
		\label{fig:corr_KIC4930889B_full_DO_highest_frequency_MD_dP}
		\hspace{0.7cm}
	\end{figure*}

	\begin{figure*}[ht!]
		\centering
		\includegraphics[width=0.7\hsize]{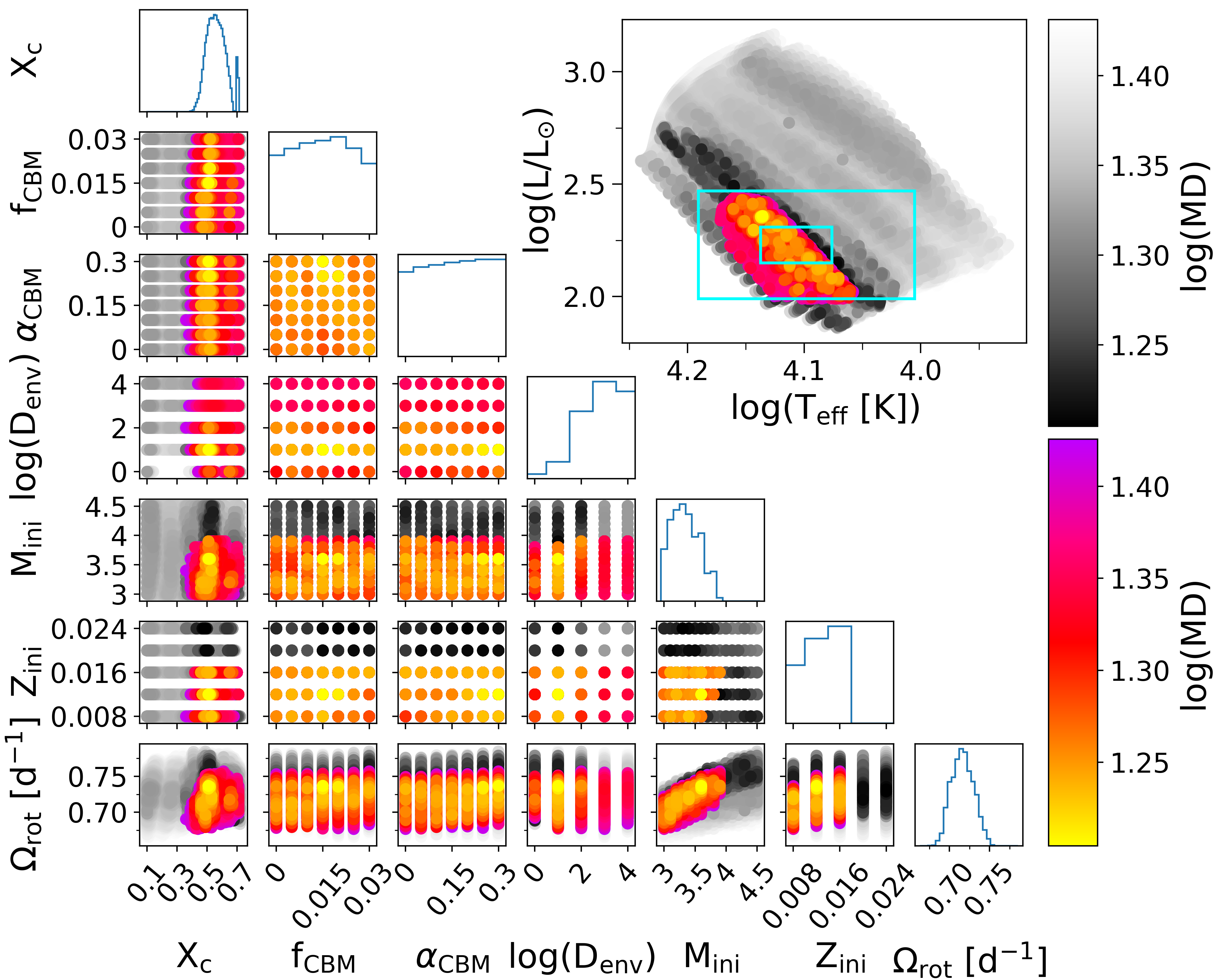}
		\caption{Corner plot (as in \cref{fig:corr_KIC4930889A_full_DO_highest_frequency_MD_dP}) for the P\'eclet grid. Made using period spacings in a Mahalanobis distance merit function and spectroscopic and luminosity constraints from the secondary star.}
		\label{fig:corr_KIC4930889B_full_ECP_highest_frequency_MD_dP}
		\hspace{0.7cm}
	\end{figure*}

	\begin{figure*}[ht!]
		\centering
		\begin{subfigure}{\hsize}
			\includegraphics[width=\hsize]{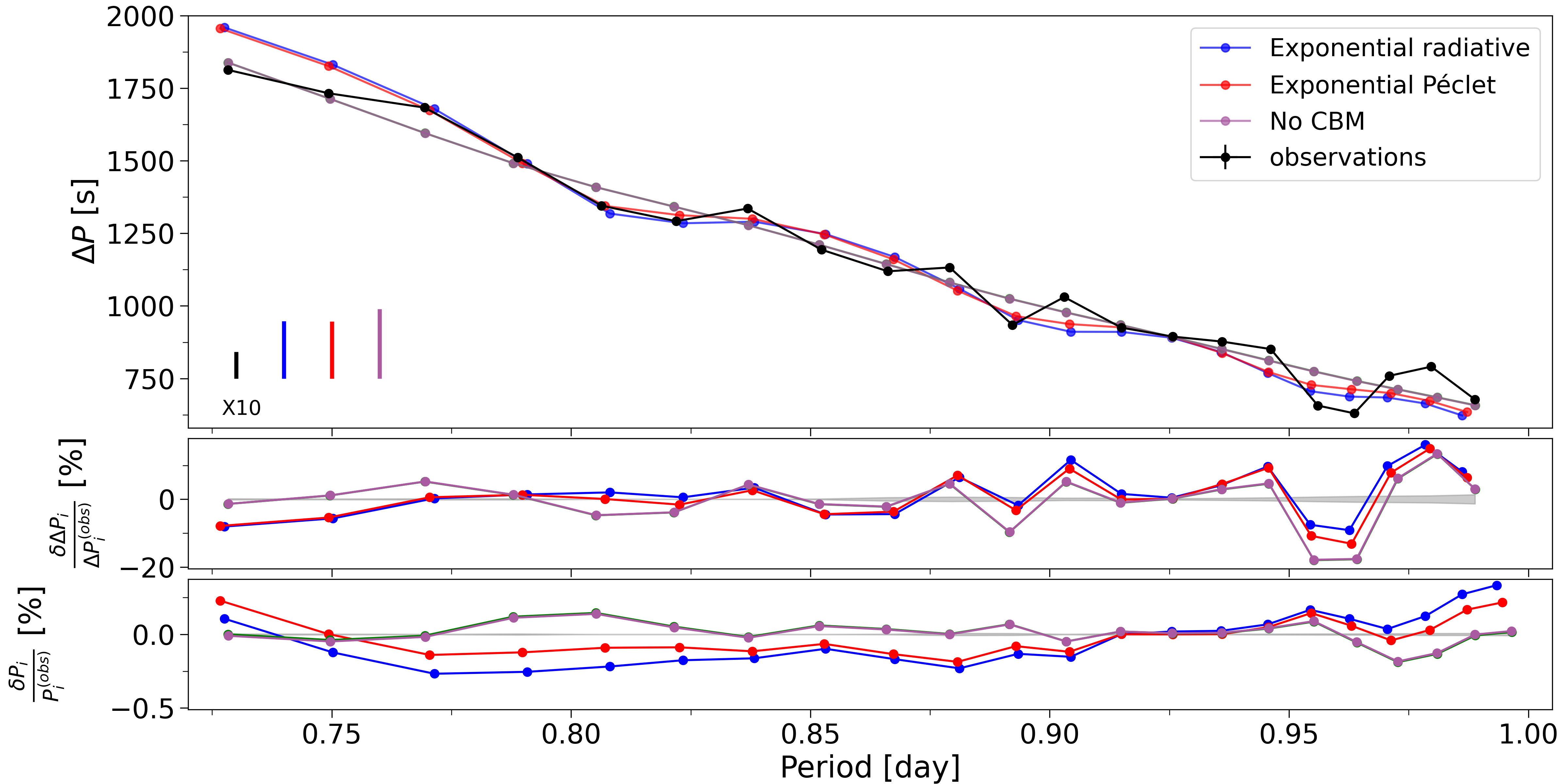}
			\caption{Period spacings for the best models for the primary.}
			\label{fig:period_spacings_primary}
		\end{subfigure}
		\begin{subfigure}{\hsize}
			\includegraphics[width=\hsize]{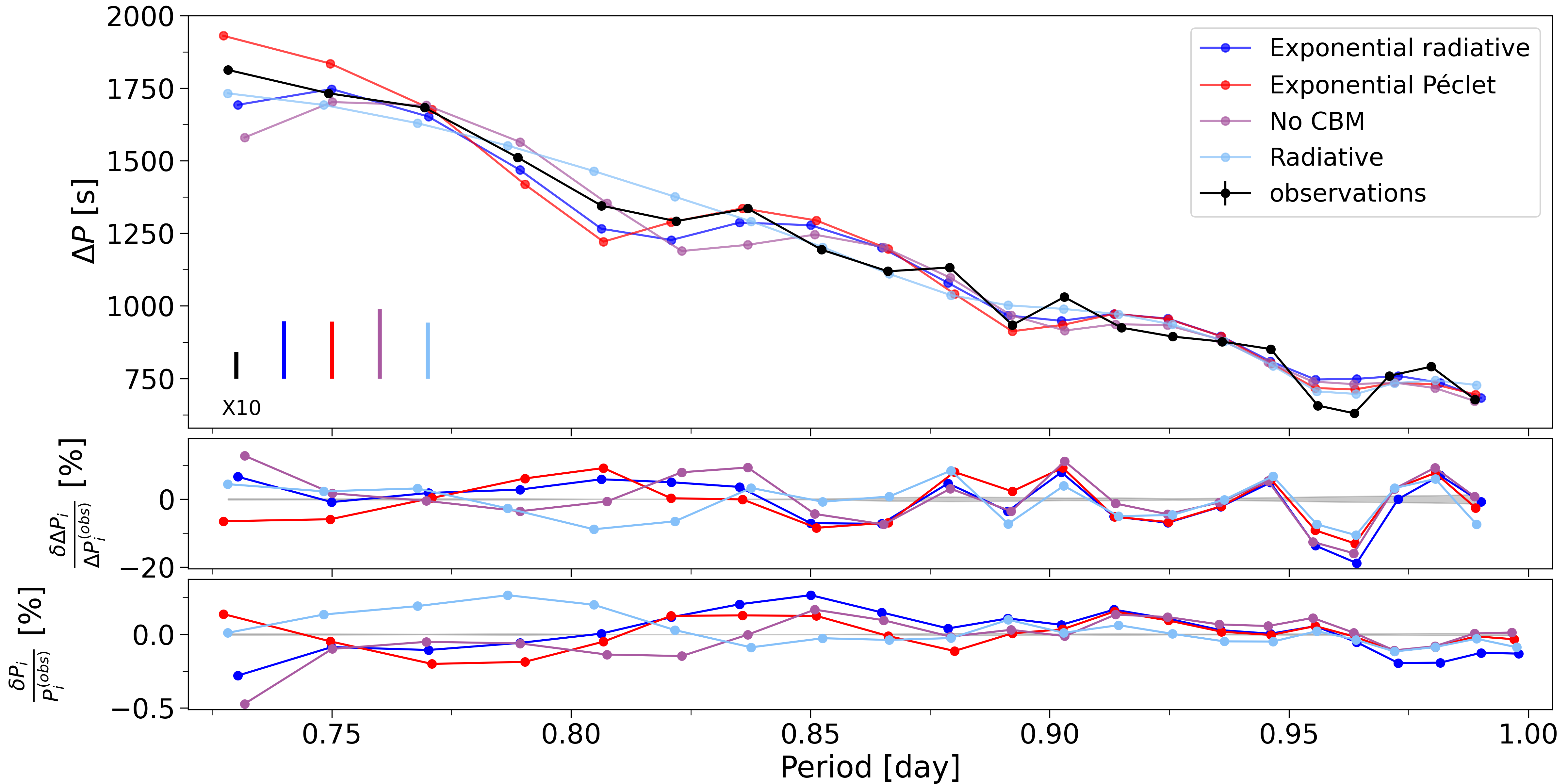}
			\caption{Period spacings for the best models for the secondary.}
			\label{fig:period_spacings_secondary}
		\end{subfigure}
		\caption{Period-spacing patterns of the observations, and of the best models of the preferred grids that are not distinguishable from one another. These are the models in bold in \cref{tab:best_models_MD_primary,tab:best_models_MD_secondary} that use $\Delta$P as observables.
			The formal errors on the observations are smaller than the symbol sizes. The largest of the observational errors is enlarged ten times and shown for comparison. The vertical bars in the bottom left corner of the top panel show the maximum considered uncertainty for the theoretical predictions approximated by the variance--covariance matrix of that particular grid.
			The middle and bottom panels show the relative difference in period spacing and period, respectively, between the observation and the model. The narrow grey areas indicate the formal 1$\sigma$ observational uncertainty from \cref{tab:freq_list}.}
		\label{fig:period_spacings}
	\end{figure*}

%%%%%%%%%%%%%%%%%%%%%%%%%%%%%%
	\begin{table*}[ht!]
	\caption{ Best-fit models of the nested grids for KIC\,4930889\,A according to Mahalanobis distance.}
	\label{tab:best_models_MD_primary}
	\centering
	\setlength\tabcolsep{3.0pt} % default value: 6pt
	\renewcommand{\arraystretch}{1.5}
	\begin{tabular}{>{\rowmac}l >{\rowmac}l >{\rowmac}l >{\rowmac}l >{\rowmac}l >{\rowmac}l >{\rowmac}l >{\rowmac}l >{\rowmac}l >{\rowmac}l >{\rowmac}l >{\rowmac}r >{\rowmac}r<{\clearrow}}
		\hline
		\hline
		Obs. &Grid  & $M_{\rm ini}$ [\msol] & $Z_{\rm ini}$ & $\acbm$ & $\fcbm$ & log($\D[env]$) & $X_{\rm c}$ & $\Omega_{\text{rot}}$ [d$^{-1}$] & $\Omega_{\text{rot}}$/$\Omega_{\text{crit}}$ & $M_{\rm cc}$[\msol] & MD & AICc \vspace{1pt} \\
		\hline

		\setrow{\bfseries} $\Delta $P & P\'eclet & 4.1${^{ 4.5 }_{ 3.3 }}$ & 0.016${^{ 0.016 }_{ 0.008 }} $ &  (...)  & 0.02${^{ 0.03 }_{ 0.00 }} $ & 2.0${^{ 4.0 }_{ 0.0 }} $ & 0.51${^{ 0.59 }_{ 0.11 }} $ & 0.75${^{ 0.76 }_{ 0.68 }} $ & 0.39 & 0.79 ${^{0.86}_{0.42}}$ & 14 & $\mathbf{-250.2}$ \\
		\setrow{\bfseries}            & Radiative  & 4.2${^{ 4.5 }_{ 3.4 }}$ & 0.016${^{ 0.016 }_{ 0.008 }} $ &  (...)  & 0.015${^{ 0.03 }_{ 0.00 }} $ & 2.0${^{ 4.0 }_{ 1.0 }} $ & 0.51${^{ 0.59 }_{ 0.11 }} $ & 0.75${^{ 0.76 }_{ 0.68 }} $ & 0.39 & 0.82 ${^{0.89}_{0.42}}$ & 13 & $\mathbf{-250.0}$ \\
		\setrow{\bfseries}            & No CBM    & 4.2${^{ 4.5 }_{ 3.6 }}$ & 0.016${^{ 0.016 }_{ 0.008 }} $ &  (...)  &  (...)  & 4.0${^{ 4.0 }_{ 2.0 }} $ & 0.42${^{ 0.57 }_{ 0.10 }} $ & 0.73${^{ 0.75 }_{ 0.67 }} $ & 0.45 & 0.72 ${^{0.85}_{0.40}}$ & 8 & $\mathbf{-249.0}$ \\
		& P\'eclet  & 4.2${^{ 4.5 }_{ 3.3 }}$ & 0.016${^{ 0.016 }_{ 0.008 }} $ & 0.15${^{ 0.3 }_{ 0.0 }} $ &  (...)  & 2.0${^{ 4.0 }_{ 1.0 }} $ & 0.50${^{ 0.58 }_{ 0.11 }} $ & 0.75${^{ 0.77 }_{ 0.68 }} $ & 0.39 & 0.79 ${^{0.86}_{0.42}}$ & 10 & $\mathbf{-248.1}$ \\
		& P\'eclet  & 3.9${^{ 4.5 }_{ 3.1 }}$ & 0.012${^{ 0.016 }_{ 0.008 }} $ & 0.3${^{ 0.3 }_{ 0.0 }} $ & 0.025${^{ 0.03 }_{ 0.00 }} $ & 1.0${^{ 4.0 }_{ 0.0 }} $ & 0.52${^{ 0.60 }_{ 0.12 }} $ & 0.75${^{ 0.78 }_{ 0.69 }} $ & 0.38 & 0.71 ${^{0.86}_{0.44}}$ & 16 & $\mathbf{-247.2}$ \\
		& Radiative & 3.9${^{ 4.5 }_{ 3.4 }}$ & 0.008${^{ 0.016 }_{ 0.008 }} $ & 0.2${^{ 0.3 }_{ 0.0 }} $ &  (...)  & 2.0${^{ 4.0 }_{ 2.0 }} $ & 0.34${^{ 0.59 }_{ 0.10 }} $ & 0.71${^{ 0.76 }_{ 0.67 }} $ & 0.44 & 0.61 ${^{0.89}_{0.40}}$ & 10 & $\mathbf{-247.0}$ \\
		& Radiative & 4.1${^{ 4.5 }_{ 3.0 }}$ & 0.016${^{ 0.016 }_{ 0.008 }} $ & 0.15${^{ 0.3 }_{ 0.0 }} $ & 0.005${^{ 0.03 }_{ 0.00 }} $ & 2.0${^{ 4.0 }_{ 0.0 }} $ & 0.52${^{ 0.60 }_{ 0.10 }} $ & 0.75${^{ 0.77 }_{ 0.67 }} $ & 0.38 & 0.81 ${^{0.91}_{0.40}}$ & 16 & $\mathbf{-244.9}$ \\

		\hline
		\setrow{\bfseries} Period & Radiative      & 4.2${^{ 4.5 }_{ 3.4 }}$ & 0.016${^{ 0.016 }_{ 0.008 }} $ &  (...)  & 0.015${^{ 0.03 }_{ 0.00 }} $ & 2.0${^{ 4.0 }_{ 1.0 }} $ & 0.51${^{ 0.59 }_{ 0.11 }} $ & 0.75${^{ 0.76 }_{ 0.68 }} $ & 0.39 & 0.82 ${^{0.89}_{0.42}}$ & 13 & $\mathbf{-269.9}$ \\
		\setrow{\bfseries}        & P\'eclet     & 4.1${^{ 4.5 }_{ 3.3 }}$ & 0.016${^{ 0.016 }_{ 0.008 }} $ &  (...)  & 0.025${^{ 0.03 }_{ 0.00 }} $ & 2.0${^{ 4.0 }_{ 1.0 }} $ & 0.52${^{ 0.58 }_{ 0.12 }} $ & 0.75${^{ 0.76 }_{ 0.69 }} $ & 0.39 & 0.80 ${^{0.86}_{0.42}}$ & 14 & $\mathbf{-269.9}$ \\
		\setrow{\bfseries}        & No CBM        & 4.2${^{ 4.5 }_{ 3.6 }}$ & 0.016${^{ 0.016 }_{ 0.008 }} $ &  (...)  &  (...)  & 4.0${^{ 4.0 }_{ 2.0 }} $ & 0.42${^{ 0.57 }_{ 0.10 }} $ & 0.73${^{ 0.75 }_{ 0.67 }} $ & 0.45 & 0.72 ${^{0.84}_{0.40}}$ & 8 & $\mathbf{-268.8}$ \\
		& P\'eclet      & 4.0${^{ 4.5 }_{ 3.3 }}$ & 0.012${^{ 0.016 }_{ 0.008 }} $ & 0.2${^{ 0.3 }_{ 0.0 }} $ &  (...)  & 2.0${^{ 4.0 }_{ 2.0 }} $ & 0.52${^{ 0.58 }_{ 0.11 }} $ & 0.75${^{ 0.77 }_{ 0.68 }} $ & 0.36 & 0.76 ${^{0.86}_{0.42}}$ & 10 & $\mathbf{-267.3}$ \\
		& Radiative    & 3.8${^{ 4.5 }_{ 3.4 }}$ & 0.008${^{ 0.016 }_{ 0.008 }} $ & 0.2${^{ 0.3 }_{ 0.0 }} $ &  (...)  & 2.0${^{ 4.0 }_{ 2.0 }} $ & 0.23${^{ 0.58 }_{ 0.10 }} $ & 0.70${^{ 0.76 }_{ 0.67 }} $ & 0.54 & 0.50 ${^{0.89}_{0.40}}$ & 10 & $\mathbf{-266.6}$ \\
		& P\'eclet      & 4.0${^{ 4.5 }_{ 3.1 }}$ & 0.012${^{ 0.016 }_{ 0.008 }} $ & 0.25${^{ 0.3 }_{ 0.0 }} $ & 0.025${^{ 0.03 }_{ 0.00 }} $ & 1.0${^{ 4.0 }_{ 0.0 }} $ & 0.52${^{ 0.60 }_{ 0.12 }} $ & 0.75${^{ 0.78 }_{ 0.69 }} $ & 0.38 & 0.75 ${^{0.86}_{0.44}}$ & 17 & $\mathbf{-266.3}$ \\
		& Radiative    & 4.1${^{ 4.5 }_{ 3.1 }}$ & 0.016${^{ 0.016 }_{ 0.008 }} $ & 0.15${^{ 0.3 }_{ 0.0 }} $ & 0.005${^{ 0.03 }_{ 0.00 }} $ & 2.0${^{ 4.0 }_{ 0.0 }} $ & 0.52${^{ 0.60 }_{ 0.10 }} $ & 0.75${^{ 0.77 }_{ 0.67 }} $ & 0.38 & 0.81 ${^{0.91}_{0.40}}$ & 16 & $\mathbf{-264.4}$ \\

		\hline
	\end{tabular}
	\begin{flushleft}
	\textbf{Notes.} The rows in bold indicate that there is no preference between this model and the best one according to the AICc. The numbers in sub- and superscript indicate the lower- and upper limits of the uncertainty region of a parameter as derived from the 2$\sigma$ error ellipses of the MD. These are projections of the error ellipses in one dimension, and do not show the parameter combinations that constitute the higher dimensional error ellipses. The method to construct the theoretical pattern is according to the highest frequency when using the period spacings as observables, and according to the highest amplitude when using the periods. Parameters fixed to zero in a nested grid are indicated by $(...)$.
	\end{flushleft}
\end{table*}

\begin{table*}[ht!]
	\caption{ Same as \cref{tab:best_models_MD_primary}, but for KIC\,4930889\,B.}
	\label{tab:best_models_MD_secondary}
	\centering
	\setlength\tabcolsep{3.0pt} % default value: 6pt
	\renewcommand{\arraystretch}{1.5}
	\begin{tabular}{>{\rowmac}l >{\rowmac}l >{\rowmac}l >{\rowmac}l >{\rowmac}l >{\rowmac}l >{\rowmac}l >{\rowmac}l >{\rowmac}l >{\rowmac}l >{\rowmac}l >{\rowmac}r >{\rowmac}r<{\clearrow}}
		\hline
		\hline
		Obs. &Grid  & $M_{\rm ini}$ [\msol] & $Z_{\rm ini}$ & $\acbm$ & $\fcbm$ & log($\D[env]$) & $X_{\rm c}$ & $\Omega_{\text{rot}}$ [d$^{-1}$] & $\Omega_{\text{rot}}$/$\Omega_{\text{crit}}$ & $M_{\rm cc}$ [\msol]& MD & AICc \vspace{1pt} \\
		\hline

		\setrow{\bfseries} $\Delta $P & P\'eclet  & 3.6${^{ 3.9 }_{ 3.0 }}$ & 0.016${^{ 0.016 }_{ 0.008 }} $ &  (...) & 0.03${^{ 0.03 }_{ 0.00 }} $ & 1.0${^{ 4.0 }_{ 0.0 }} $ & 0.52${^{ 0.71 }_{ 0.36 }} $ & 0.73${^{ 0.75 }_{ 0.67 }} $ & 0.37 & 0.67 ${^{0.84}_{0.47}}$ & 13 & $\mathbf{-251.0}$ \\
		\setrow{\bfseries}            & Radiative & 3.2${^{ 3.9 }_{ 3.0 }}$ & 0.012${^{ 0.016 }_{ 0.008 }} $ &  (...)  & 0.03${^{ 0.03 }_{ 0.00 }} $ & 1.0${^{ 4.0 }_{ 0.0 }} $ & 0.52${^{ 0.71 }_{ 0.33 }} $ & 0.71${^{ 0.75 }_{ 0.67 }} $ & 0.33 & 0.60 ${^{0.86}_{0.45}}$ & 12 & $\mathbf{-250.8}$ \\
		\setrow{\bfseries}            & Radiative & 3.0${^{ 3.9 }_{ 3.0 }}$ & 0.008${^{ 0.016 }_{ 0.008 }} $ & 0.25${^{ 0.3 }_{ 0.0 }} $ & 0.01${^{ 0.03 }_{ 0.00 }} $ & 1.0${^{ 4.0 }_{ 0.0 }} $ & 0.62${^{ 0.71 }_{ 0.33 }} $ & 0.69${^{ 0.75 }_{ 0.67 }} $ & 0.24 & 0.61 ${^{0.88}_{0.44}}$ & 10 & $\mathbf{-250.1}$ \\
		\setrow{\bfseries}            & No CBM     & 3.8${^{ 3.9 }_{ 3.0 }}$ & 0.012${^{ 0.016 }_{ 0.008 }} $ &  (...)  &  (...)  & 2.0${^{ 4.0 }_{ 0.0 }} $ & 0.49${^{ 0.71 }_{ 0.33 }} $ & 0.73${^{ 0.74 }_{ 0.67 }} $ & 0.35 & 0.67 ${^{0.83}_{0.44}}$ & 8 & $\mathbf{-249.4}$ \\
		& Radiative & 3.0${^{ 3.9 }_{ 3.0 }}$ & 0.016${^{ 0.016 }_{ 0.008 }} $ & 0.2${^{ 0.3 }_{ 0.0 }} $ &  (...)  & 1.0${^{ 4.0 }_{ 0.0 }} $ & 0.51${^{ 0.71 }_{ 0.33 }} $ & 0.69${^{ 0.75 }_{ 0.67 }} $ & 0.32 & 0.53 ${^{0.86}_{0.44}}$ & 8 & $\mathbf{-248.4}$ \\
		& P\'eclet  & 3.9${^{ 3.9 }_{ 3.0 }}$ & 0.016${^{ 0.016 }_{ 0.008 }} $ & 0.05${^{ 0.3 }_{ 0.0 }} $ &  (...)  & 2.0${^{ 4.0 }_{ 0.0 }} $ & 0.51${^{ 0.71 }_{ 0.36 }} $ & 0.74${^{ 0.75 }_{ 0.67 }} $ & 0.36 & 0.72 ${^{0.84}_{0.46}}$ & 10 & $\mathbf{-248.4}$ \\
		& P\'eclet  & 3.6${^{ 3.9 }_{ 3.0 }}$ & 0.012${^{ 0.016 }_{ 0.008 }} $ & 0.3${^{ 0.3 }_{ 0.0 }} $ & 0.015${^{ 0.03 }_{ 0.00 }} $ & 1.0${^{ 4.0 }_{ 0.0 }} $ & 0.51${^{ 0.71 }_{ 0.36 }} $ & 0.74${^{ 0.76 }_{ 0.68 }} $ & 0.37 & 0.64 ${^{0.84}_{0.46}}$ & 16 & $\mathbf{-247.6}$ \\

		\hline

		\setrow{\bfseries} Period & Radiative     & 3.2${^{ 3.9 }_{ 3.0 }}$ & 0.016${^{ 0.016 }_{ 0.008 }} $ &  (...)  & 0.03${^{ 0.03 }_{ 0.00 }} $ & 1.0${^{ 4.0 }_{ 0.0 }} $ & 0.51${^{ 0.71 }_{ 0.33 }} $ & 0.72${^{ 0.75 }_{ 0.67 }} $ & 0.36 & 0.59 ${^{0.86}_{0.45}}$ & 12 & $\mathbf{-270.7}$ \\
		\setrow{\bfseries}        & P\'eclet      & 3.5${^{ 3.9 }_{ 3.0 }}$ & 0.016${^{ 0.016 }_{ 0.008 }} $ &  (...)  & 0.03${^{ 0.03 }_{ 0.00 }} $ & 1.0${^{ 4.0 }_{ 0.0 }} $ & 0.52${^{ 0.71 }_{ 0.36 }} $ & 0.72${^{ 0.75 }_{ 0.67 }} $ & 0.36 & 0.65 ${^{0.84}_{0.47}}$ & 13 & $\mathbf{-270.5}$ \\
		\setrow{\bfseries}        & Radiative     & 3.0${^{ 3.9 }_{ 3.0 }}$ & 0.008${^{ 0.016 }_{ 0.008 }} $ & 0.25${^{ 0.3 }_{ 0.0 }} $ & 0.01${^{ 0.03 }_{ 0.00 }} $ & 1.0${^{ 4.0 }_{ 0.0 }} $ & 0.62${^{ 0.71 }_{ 0.33 }} $ & 0.69${^{ 0.76 }_{ 0.66 }} $ & 0.24 & 0.61 ${^{0.88}_{0.44}}$ & 10 & $\mathbf{-269.9}$ \\
		\setrow{\bfseries}        & No CBM        & 3.7${^{ 3.9 }_{ 3.0 }}$ & 0.012${^{ 0.016 }_{ 0.008 }} $ &  (...)  &  (...)  & 2.0${^{ 4.0 }_{ 0.0 }} $ & 0.49${^{ 0.71 }_{ 0.33 }} $ & 0.72${^{ 0.74 }_{ 0.67 }} $ & 0.34 & 0.65 ${^{0.83}_{0.44}}$ & 8 & $\mathbf{-269.2}$ \\
		& Radiative     & 3.0${^{ 3.9 }_{ 3.0 }}$ & 0.016${^{ 0.016 }_{ 0.008 }} $ & 0.2${^{ 0.3 }_{ 0.0 }} $ &  (...)  & 1.0${^{ 4.0 }_{ 0.0 }} $ & 0.51${^{ 0.71 }_{ 0.33 }} $ & 0.69${^{ 0.75 }_{ 0.66 }} $ & 0.32 & 0.53 ${^{0.86}_{0.44}}$ & 8 & $\mathbf{-268.3}$ \\
		& P\'eclet      & 3.8${^{ 3.9 }_{ 3.0 }}$ & 0.016${^{ 0.016 }_{ 0.008 }} $ & 0.1${^{ 0.3 }_{ 0.0 }} $ &  (...)  & 2.0${^{ 4.0 }_{ 0.0 }} $ & 0.51${^{ 0.71 }_{ 0.35 }} $ & 0.73${^{ 0.75 }_{ 0.67 }} $ & 0.36 & 0.70 ${^{0.84}_{0.46}}$ & 10 & $\mathbf{-267.6}$ \\
		& P\'eclet      & 3.6${^{ 3.9 }_{ 3.0 }}$ & 0.012${^{ 0.016 }_{ 0.008 }} $ & 0.3${^{ 0.3 }_{ 0.0 }} $ & 0.02${^{ 0.03 }_{ 0.00 }} $ & 1.0${^{ 4.0 }_{ 0.0 }} $ & 0.52${^{ 0.71 }_{ 0.36 }} $ & 0.74${^{ 0.76 }_{ 0.67 }} $ & 0.36 & 0.65 ${^{0.84}_{0.46}}$ & 17 & $\mathbf{-266.5}$ \\

		\hline
	\end{tabular}
\end{table*}

	The condition numbers of the variance-covariance matrices
	computed via Eq.\,(\ref{eq:cond_nr}) are of order $\kappa(A) \sim 10^3$ to $10^4$ when considering mode periods, but are significantly smaller, down to
	$\kappa(A) \sim 10^1$, when considering period spacings. We therefore primarily consider the period spacings as the set of observables to fit, but still list the results from using the periods as observables as well. Although the individual parameters of the best models may differ between these two sets of observables, they are in most cases quite similar if not the same, and always fall within the other error ellipse. Our conclusions are therefore independent from the chosen observable.

	We model the observed pulsation pattern twice. Once with the spectroscopic constraints of the primary, for which the corner plots of the radiative and P\'eclet grid are shown in \cref{fig:corr_KIC4930889A_full_DO_highest_frequency_MD_dP,fig:corr_KIC4930889A_full_ECP_highest_frequency_MD_dP}, and once with the spectroscopic constraints of the secondary, with \cref{fig:corr_KIC4930889B_full_DO_highest_frequency_MD_dP,fig:corr_KIC4930889B_full_ECP_highest_frequency_MD_dP} showing the corner plots for the radiative and P\'eclet grid respectively. The models included in the 2$\sigma$ error ellipse of the MD according to \cref{eq:error_ellips} are shown in colour, while the models in grey scale fall outside of this error ellipse.
	Additionally we make a comparison between the AICc values of the best models of the full grid and each partial grid with fewer free parameters, for both prescriptions of the temperature gradient in the CBM region. We hereafter refer to the grids with six free parameters as the radiative and P\'eclet grid, but specify when talking about grids with fewer free CBM parameters. Among the nested grids with five free parameters, we have $\acbm=0$ but varying $\fcbm$, henceforth denoted as exponential radiative or exponential P\'eclet grid, and $\fcbm=0$ but varying $\acbm$, henceforth denoted as step radiative or step P\'eclet grid. The nested grid with four free parameters, having both $\acbm$ and $\fcbm$ set to zero, is referred to as the grid without CBM.
	Comparing all these grids with various numbers of free parameters not only enables us to investigate which temperature gradient is preferred in the CBM region, but also to examine if the increased fit quality outweighs the penalties for higher model complexity.

	The model parameters and AICc values of these best models are listed in \cref{tab:best_models_MD_primary} and \cref{tab:best_models_MD_secondary} when enforcing the constraints on the luminosity and spectroscopic $\Teff$ and $\log\,g$ of the primary and secondary, respectively.
	We cannot distinguish the preferred temperature gradient whilst modelling the primary star, since there is no preference between the exponential radiative, exponential P\'eclet, or the grid without any CBM, given that $\Delta \text{AICc} < 2$ between their best models. We do however find a preference of these three grids over the grids with a step-like mixing profile, or the ones with a combined step and exponential mixing. The period spacings of the best models from these indistinguishable grids are shown in \cref{fig:period_spacings_primary}.

	Modelling the secondary star also yields no possibility to distinguish which temperature gradient is preferred. It is in this case not possible to differentiate between both grids with an exponentially decaying mixing in the CBM region, the grid without any CBM, and the radiative grid with six free parameters. The period spacings of the best models from these indistinguishable grids are shown in \cref{fig:period_spacings_secondary}.

	We clearly see from both \cref{fig:period_spacings_primary} and \cref{fig:period_spacings_secondary} that the variance of the theoretical predictions
	is much larger than the uncertainties on the observations. Our solutions are therefore dominated by the theoretical uncertainties, rather than the observational ones. The larger variance of the grid without CBM is one of the reasons why there is no selection capacity between this grid and those with exponential CBM. A reduction of the variance of the theoretical predictions would lead to a much stronger preference for the presence of CBM over the absence of CBM according to the AICc.
	This would in particular be the case should we ignore the (co)variance due to limits in the theoretical predictions ($V$=0), that is when reducing the merit function from a Mahalanobis distance to a $\chi^2$.

	\subsection{Near-core rotation rate and convective core mass}

	From the 2$\sigma$ MD error ellipses on the best models, we find the near-core rotation rate of the star to be well constrained. The values are listed alongside the best models of each nested grid in \cref{tab:best_models_MD_primary,tab:best_models_MD_secondary}. If we take the grids that are indistinguishable according to the AICc, we find all of them to be consistent with the result of our best model grid; $\Omega_{\text{rot}}=0.73^{+0.02}_{-0.06} \mathrm{d}^{-1}$. This is about 13.4 times the orbital frequency of this eccentric binary, and about 37\% of the best model's Roche critical rotation rate.
	Even when considering the nested grids that are not preferred, we can see that their near-core rotation rates are also consistent and agree very well with one another.

	Defining the convective core mass as the one determined by the Ledoux criterion
	without including the CBM region (as is visualised by the grey area in \cref{fig:mixing_profile}), we constrain it to $M_{\rm cc} =0.67^{+0.17}_{-0.20}$\,M$_\odot$. This value is consistent across all grids we considered in the modelling, keeping in mind its uncertainties.

	\subsection{Mode excitation}

	We are left with multiple different solutions that cannot be distinguished from each other based on modelling the period-spacing values and using the spectroscopic, astrometric, and isochrone-cloud restraints. Therefore we look at which of these models performs best at reproducing the mode excitation of our observed period-spacing pattern.

\begin{figure*}[ht]
	\begin{minipage}[c]{0.8\textwidth}
		\includegraphics[width=\hsize]{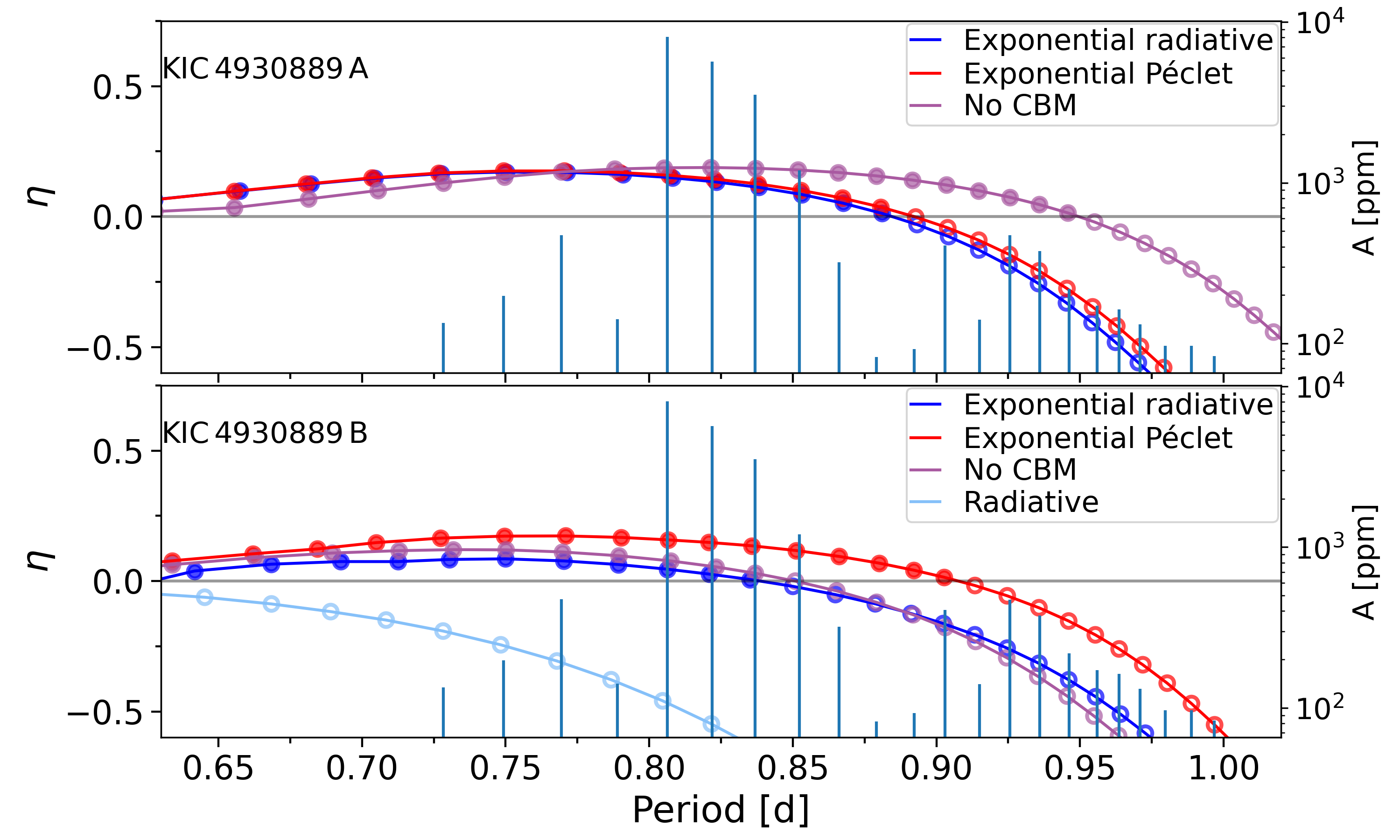}
	\end{minipage}\hfill
	\begin{minipage}[c]{0.19\textwidth}
		\caption{Instability parameter $\eta$. The parameter is shown as a function of mode period for the best models of the primary (top panel) and secondary (bottom panel). The period spacing patterns of these models are shown in \cref{fig:period_spacings}, coloured circles indicate excited modes, while empty circles indicate the non-excited ones. The vertical lines show the observed mode periods and their amplitude.}
		\label{fig:mode_excitation}
		\end{minipage}
	\end{figure*}

	\cref{fig:mode_excitation} shows the normalised growth rates, $\eta$, of the modes \citep{1978AJ.....83.1184S}. These indicate an excited or damped mode for a positive or negative value of $\eta$, respectively.
	For the primary, the models from the exponential P\'eclet and the exponential radiative grid have ten modes excited out of the 22 observed modes in our pulsation pattern. The model from the grid without CBM has sixteen of the observed modes as excited. This higher amount of excited modes is an effect of its more evolved nature compared to the other best models, rather than a direct effect of the absence of CBM \citep[e.g. Fig. 1. of][which shows an increasing number excited modes during the first part of the main-sequence evolution]{2017A&A...598A..74P}.
	All of these models show some excited modes at shorter periods that were not observed in our pattern.
	The model for the secondary star from the exponential P\'eclet grid shows twelve out of the 22 modes excited. The models from the exponential radiative grid and the grid without CBM show seven excited modes, while the model from the radiative grid shows no excited modes at all.

	Accurately reproducing the excitation of high radial order g modes in B-type pulsators often requires opacity enhancements which were not considered in this study \citep[e.g.][]{2016MNRAS.455L..67M,2017MNRAS.466.2284D,2019MNRAS.485.3544W,2022MNRAS.511.1529S}.
	In our results, we also see that the increased number of excited modes in general corresponds to an increased metallicity, and hence elevated opacity of the iron- and nickel-group chemical elements. We therefore confirm that the standard OP opacity tables are insufficient to accurately reproduce the observed mode excitations, rather than using this result to constrain our solutions.

%	\begin{figure*}[ht]
%		\centering
%		\includegraphics[width=0.9\hsize]{figures/mode_excitation.png}
%		\caption{Instability parameter $\eta$. The parameter is shown as a function of mode period for the best models of the primary (top panel) and secondary (bottom panel). The period spacing patterns of these models are shown in \cref{fig:period_spacings}, coloured circles indicate excited modes, while empty circles indicate the non-excited ones. The vertical lines show the observed mode periods and their amplitude.}
%		\label{fig:mode_excitation}
%	\end{figure*}

	%%%%%%%%%%%%%%%%%%%%%%%%%%%%%%%%%%%%%%%%%%%%%%%%%%%%%%%%%%%%%%%%%%%%%%%%%%%%%%%%%%%%%%%%%%%%%%
	\section{Conclusions}\label{sec:conclusions}

	In this work we investigated the gravito-inertial modes in the double-lined B-type binary KIC\,4930889. We explored which of the components hosts the pulsations, what the preferred temperature gradient and mixing profiles are in the CBM region of that star, and constrained its near-core rotation rate.
	We employed asteroseismic, spectroscopic and astrometric information, and constraints obtained from the binarity through isochrone clouds.

	The quality of our best asteroseismic solutions are better for the spectroscopic and astrometric constraints of the secondary star, although not statistically significantly better than when using those of the primary star. The difference in luminosity between both stars, with 67\% and 33\% of the light contribution from the primary and secondary respectively, is not large enough to assign the pulsation signal to one or the other.
	Furthermore, we do not find a preference for the temperature gradient based on the 22 mode periods or 21 period spacings. However, we are able to constrain the near-core rotation rate of the pulsating component to $\Omega_{\text{rot}}=0.73^{+0.02}_{-0.06} \mathrm{d}^{-1}$.
	We also obtain better solutions for models with some type of exponentially decaying CBM over those with a step-like mixing profile in the CBM region.

	We find that the model from the exponential P\'eclet grid performs better at explaining the mode excitation than the models with a radiative temperature gradient. The P\'eclet model shows twelve out of the 22 observed modes excited, whereas the radiative models have at most seven of the observed modes excited.
	The larger number of excited modes is due to the higher metallicity of the stellar equilibrium model selected from this grid; that is, the different temperature gradient only indirectly influences the predicted mode excitation. The model that best explains the excited modes is however the one without CBM present that assumes the spectroscopic and astrometric constraints for the primary star. This absence of CBM also only indirectly influences the mode excitation, since this model is more evolved than the others along the first half of its main sequence, entailing a higher number of theoretically predicted excited modes.

	Comparing our detailed follow-up treatment to the earlier performed statistical modelling approach of \citet{2021NatAs...5..715P} and \citet{2022ApJ...930...94P}, we note a few key differences in the modelling setup. Whilst \citet{2021NatAs...5..715P} investigated two different prescriptions for the CBM region and four for the envelope mixing, our study considers only one case of envelope mixing but seven different ones for the CBM region. We opted for such an approach since the g modes that we are considering have a much higher probing power in the CBM region than in the stellar envelope, as can be seen from their mode inertia in \cref{fig:mixing_profile}.
	\citet{2022ApJ...930...94P} was able to distinguish between different shapes of the envelope mixing, but found no preference between their considered CBM prescriptions. This would equal a comparison between our exponential radiative and step P\'eclet grids, where we do find a preference for the exponential grids over the step grids.

	As far as the retrieved model parameters are concerned, we list both the results from \citet{2022ApJ...930...94P} and from this work (using the spectroscopic and astrometric constraints of the primary and secondary) in \cref{tab:result_comparison}.
	All parameters are in agreement when considering our error estimation and considering the spectroscopic and astrometric constraints for the primary star. In particular, the stellar rotation rate aligns well. However, our best asteroseismic model was found when we consider the constraints for the secondary star. For this case the mass and central hydrogen content are no longer compatible within the projected error ellipses.
	Our best point estimators do deliver a younger, less massive star with more CBM and less envelope mixing.
	We note that the uncertainties on the two sets of results differ substantially, where our uncertainties on the parameters encompass the results from \citet{2022ApJ...930...94P}, but not vice versa.
	These different results are influenced slightly by the different spectroscopic constraints and set of prograde dipole modes that are employed, but stem dominantly from the modelling approach, where we used a more detailed treatment of the asteroseismic modelling as compared to the approximative statistical modelling approach of \citet{2021NatAs...5..715P,2022ApJ...930...94P,2022ApJ...940...49P}, who used statistical approximations for the pulsations rather than detailed \texttt{GYRE} computations. We find their uncertainties obtained from the approximative statistical modelling to be underestimated for this star.

	\begin{table}[ht!]
		\caption{Stellar parameters of KIC\,4930889 derived by \citet{2022ApJ...930...94P} and \citet{2022ApJ...940...49P} compared to the parameters from our best model. }
		\label{tab:result_comparison}
		\centering
		\renewcommand{\arraystretch}{1.15}
		\setlength\tabcolsep{4.0pt} % default value: 6pt
		\begin{tabular}{l c c c }
			\hline
			\hline
			& Primary & Primary & Secondary \\
			Parameter & \citet{2022ApJ...930...94P} & This work & This work \\
			\hline
			$M_{\rm ini}$ [\msol] &  4.06$\pm$0.31 &  4.1$^{+0.4}_{-0.8}$ &  3.6$^{+0.3}_{-0.6}$ \\
			$Z_{\rm ini}$ &  0.00924$\pm$0.00002 &  0.016$^{+0}_{-0.008}$ &  0.016$^{+0}_{-0.008}$ \\
			$\fcbm$  & 0.012$\pm$0.001 &  0.02 $^{+0.01}_{-0.02}$ &  0.03 $^{+0}_{-0.03}$ \\
			log($\D[env]$) &  3.3$\pm$0.5 &  2$^{+2}_{-2}$  &  1$^{+3}_{-1}$ \\
			$X_{\rm c}/X_{\rm ini}$ &  0.362$\pm$0.0007 &  0.72$^{+0.11}_{-0.57}$ &  0.74$^{+0.26}_{-0.23}$ \\
			\hline
			& \citet{2022ApJ...940...49P} & This work & This work \\
			\hline
			$\Omega_{\rm rot}$ (d$^{-1}$) &  0.740$\pm$0.008 & 0.75$^{+0.01}_{-0.07}$ & 0.73$^{+0.02}_{-0.06}$ \\
			\hline
		\end{tabular}
		\begin{flushleft}
			\textbf{Notes.} From top to bottom, the parameters are  initial mass, metallicity, exponential CBM parameter, envelope mixing at the CBM interface (constant for \citet{2022ApJ...930...94P}, internal gravity wave profile for this work), central hydrogen content, and near-core rotation rate. The uncertainties stem from the projections of the error ellipse on one dimension.
		\end{flushleft}
	\end{table}

	Although the error ellipses of our solutions contain less than 3\% of our initial models, the projections of the six dimensional error ellipse in one dimension results in uncertainties on each individual parameter that range over most of the initial model grid. The vast majority of the parameter combinations that are included in these one-dimensional projections are however not part of the actual higher dimensional error ellipse. An indication of this can be seen in the two-dimensional projections of the error ellipse in \cref{fig:corr_KIC4930889A_full_DO_highest_frequency_MD_dP,fig:corr_KIC4930889A_full_ECP_highest_frequency_MD_dP,fig:corr_KIC4930889B_full_DO_highest_frequency_MD_dP,fig:corr_KIC4930889B_full_ECP_highest_frequency_MD_dP}.
	In contrast, the results obtained by \citet{2021A&A...650A.175M} for KIC\,7760680 yielded much smaller error ellipses so that the uncertainties remained small even when they were projected on one dimension. This difference is due to the number of modes in the observed prograde dipole mode pattern, which amounted to 36 modes for KIC\,7760680 and to 22 modes for KIC\,4930889, where a larger number of observed modes entails a better probing power of the stellar interior.
	With this in mind, constraints from spectroscopic data, \textit{Gaia} astrometric data, and from the binarity of the system are valuable to complement asteroseismic information. These complementary constraints become all the more beneficial for stars with lower asteroseismic probing power due to fewer observed modes.

	Similar to \citet{2021A&A...650A.175M}, so regardless of the amount of pulsations in our observed period pattern, we find that the uncertainties on the theoretically predicted pulsation patterns are much larger than the uncertainties on the observed patterns. The uncertainties in our modelling are therefore dominated by the theoretical model uncertainties, rather than the observational ones. Additionally, the theoretical variance is largest in the grid without CBM, causing the lack of selection capacity between these models and the ones with CBM. A reduction of the variances would lead to a stronger preference of model grids with CBM over the ones without it. Hence, future work should prioritise improving stellar evolutionary models by both refining and expanding the physical processes that are included in them.
	KIC\,4930889 is a good target to evaluate tidal effects in close binary evolution models, given our detection of multiples of the orbital frequency in its secondary period spacing patterns.

	%%%%%%%%%%%%%%%%%%%%%%%%%%%%%%%%%%%%%%%%%%%%%%%%%%%%%%%%%%%%%%%%%%%%%%%%%%%%%%%%%%%%%%%%%%%%%%
	\begin{acknowledgements}
		The authors thank Jordan van Beeck for having provided the frequencies he found for KIC\,4930889 from his 2021 paper in electronic format, and Sarah Gebruers and Alex Kemp for the insightful scientific discussions.
		The authors are grateful to the \texttt{MESA} and \texttt{GYRE} developers teams for their efforts and for releasing their software publicly; this study would  not have been possible without their codes.
		The research leading to these results has received funding from the Research Foundation Flanders (FWO) by means of a PhD scholarship to MM under project No. 11F7120N, a postdoctoral fellowship to TVR with grant agreement No. 12ZB620N, and to AT through grant agreement No. G089422N, from the KU\,Leuven Research Council (grant C16/18/005: PARADISE to PI Aerts), as well as from the BELgian federal Science Policy Office (BELSPO) through PRODEX grant PLATO.

	\end{acknowledgements}
	%%%%%%%%%%%%%%%%%%%%%%%%%%%%%%%%%%%%%%%%%%%%%%%%%%%%%%%%%%%%%%%%%%%%%%%%%%%%%%%%%%%%%%%%%%%%%%
	\bibliographystyle{aa}

	%%%%%%%%%%%%%%%%%%%%%%%%%%%%%%%%%%%%%%%%%%%%%%%%%%%%%%%%%%%%%%%%%%%%%%%%%%%%%%%%%%%%%%%%%%%%%%
	%%%%%%%%%%%%%%%%%%%%%%%%%%%%%%%%%%%%%%%%%%%%%%%%%%%%%%%%%%%%%%%%%%%%%%%%%%%%%%%%%%%%%%%%%%%%%%
	\begin{appendix}

	\section{MESA and GYRE Inlists} \label{appendix:inlist}
	The example \texttt{MESA} and \texttt{GYRE} inlists used for this work are
	available from the \texttt{MESA} inlists section of the \texttt{MESA} marketplace:
	\url{https://cococubed.com/mesa_market/inlists.html}.

	\section{Frequency list}
	We provide lists of frequencies.
	
	\onecolumn
	\begin{center}
	\newcolumntype{d}[1]{D{.}{.}{#1}}
	\longtab[0](
	\begin{longtable}{r l l r d{3.5}}
		\caption{Full list of periods, frequencies, amplitudes, and phases of all the modes as extracted by \citet{2021A&A...655A..59V} from the \textit{Kepler} light curve. Numbers in parenthesis are the errors on the last significant digit.}
		\label{tab:complete_freq_list} \\
		\hhline{=====}
		\setlength\tabcolsep{6.0pt} % default value: 6pt
		\renewcommand{\arraystretch}{1.13}
		\# & $p$ [d] & $f$ [d$^{-1}$] & $A$ [ppm] & \multicolumn{1}{c} {$\theta \, [\rm {rad}]$} \\
		\hline
		\endfirsthead
		\caption{continued}\\
		\hhline{=====}
		\# & $p$ [d] & $f$ [d$^{-1}$] & $A$ [ppm] & \multicolumn{1}{c} {$\theta \, [\rm {rad}]$} \\
		\hline
		\endhead
		\hline
		\endfoot
		\hline
		\endlastfoot
			
		1 & 0.205512(5) & 4.86589(12) & 56(18) & 0.4(3) \\ 
		2 & 0.387072(9) & 2.58350(6) & 110(18) & -0.15(16) \\ 
		3 & 0.39372(2) & 2.53989(13) & 51(18) & -0.7(4) \\ 
		4 & 0.39597(2) & 2.52547(14) & 48(18) & -0.8(4) \\ 
		5 & 0.397390(14) & 2.51642(9) & 74(18) & 3.0(2) \\ 
		6 & 0.403177(4) & 2.48030(2) & 290(18) & -2.84(6) \\ 
		7 & 0.407034(6) & 2.45680(3) & 200(18) & 0.02(9) \\ 
		8 & 0.410673(7) & 2.43503(4) & 171(18) & -1.61(11) \\ 
		9 & 0.414671(12) & 2.41155(7) & 99(18) & 0.91(18) \\ 
		10 & 0.417729(14) & 2.39390(8) & 87(18) & -1.6(2) \\ 
		11 & 0.41844(3) & 2.38985(15) & 44(18) & -0.1(4) \\ 
		12 & 0.44826(3) & 2.23087(14) & 47(18) & 2.6(4) \\ 
		13 & 0.582174(4) & 1.717699(12) & 573(18) & 2.09(3) \\ 
		14 & 0.63246(3) & 1.58113(8) & 88(18) & -0.9(2) \\ 
		15 & 0.64199(5) & 1.55767(11) & 61(18) & -2.1(3) \\ 
		16 & 0.72834(3) & 1.37298(5) & 134(18) & 2.02(13) \\ 
		17 & 0.74442(4) & 1.34332(7) & 97(18) & 0.12(19) \\ 
		18 & 0.749329(19) & 1.33453(3) & 198(18) & -0.14(9) \\ 
		19 & 0.76795(5) & 1.30217(8) & 82(18) & 0.6(2) \\ 
		20 & 0.769383(8) & 1.299742(14) & 473(18) & -1.35(4) \\ 
		21 & 0.77807(5) & 1.28522(8) & 82(18) & -2.6(2) \\ 
		22 & 0.78345(4) & 1.27640(7) & 100(18) & 2.60(18) \\ 
		23 & 0.78887(3) & 1.26764(5) & 142(18) & -2.07(13) \\ 
		24 & 0.79136(6) & 1.26364(9) & 72(18) & -2.6(2) \\ 
		25 & 0.79654(10) & 1.25543(15) & 44(18) & 0.0(4) \\ 
		26 & 0.80074(10) & 1.24884(15) & 45(18) & -0.6(4) \\ 
		27 & 0.80209(7) & 1.24674(11) & 59(18) & -1.3(3) \\ 
		28 & 0.80392(5) & 1.24390(7) & 96(18) & 0.02(19) \\ 
		29 & 0.8063568(5) & 1.2401457(8) & 8064(18) & 2.855(2) \\ 
		30 & 0.80868(3) & 1.23659(5) & 141(18) & 1.29(13) \\ 
		31 & 0.81027(3) & 1.23416(5) & 149(18) & 1.00(12) \\ 
		32 & 0.81183(3) & 1.23179(5) & 136(18) & 1.01(13) \\ 
		33 & 0.81450(8) & 1.22775(12) & 57(18) & 0.5(3) \\ 
		34 & 0.81610(5) & 1.22533(7) & 99(18) & 0.37(18) \\ 
		35 & 0.81755(5) & 1.22316(7) & 93(18) & -0.30(19) \\ 
		36 & 0.81923(3) & 1.22066(5) & 144(18) & -0.28(13) \\ 
		37 & 0.8219273(8) & 1.2166526(12) & 5677(18) & -1.942(3) \\ 
		38 & 0.82348(3) & 1.21436(4) & 154(18) & -2.40(12) \\ 
		39 & 0.82545(4) & 1.21146(6) & 108(18) & -0.03(17) \\ 
		40 & 0.82688(4) & 1.20936(6) & 106(18) & -1.15(17) \\ 
		41 & 0.82995(5) & 1.20489(8) & 87(18) & -2.0(2) \\ 
		42 & 0.83275(4) & 1.20083(6) & 120(18) & -3.09(15) \\ 
		43 & 0.8368807(13) & 1.1949135(19) & 3529(18) & -2.774(5) \\ 
		44 & 0.83867(3) & 1.19236(4) & 170(18) & -2.34(11) \\ 
		45 & 0.84108(7) & 1.18895(9) & 73(18) & -2.2(2) \\ 
		46 & 0.84350(8) & 1.18553(11) & 61(18) & 2.8(3) \\ 
		47 & 0.84604(8) & 1.18197(11) & 62(18) & -0.4(3) \\ 
		48 & 0.84838(7) & 1.17872(10) & 70(18) & -2.5(3) \\ 
		49 & 0.852339(4) & 1.173242(6) & 1197(18) & -1.522(15) \\ 
		50 & 0.853630(7) & 1.171467(10) & 686(18) & -1.62(3) \\ 
		51 & 0.85532(7) & 1.16916(10) & 66(18) & -2.3(3) \\ 
		52 & 0.85997(8) & 1.16283(10) & 64(18) & 3.0(3) \\ 
		53 & 0.86475(9) & 1.15640(12) & 54(18) & -1.6(3) \\ 
		54 & 0.866155(16) & 1.15453(2) & 320(18) & 3.04(6) \\ 
		55 & 0.86905(5) & 1.15068(7) & 100(18) & -2.61(18) \\ 
		56 & 0.87152(5) & 1.14743(6) & 111(18) & 2.73(16) \\ 
		57 & 0.87608(3) & 1.14145(4) & 180(18) & 1.98(10) \\ 
		58 & 0.87911(6) & 1.13752(8) & 82(18) & -2.5(2) \\ 
		59 & 0.88242(10) & 1.13325(13) & 54(18) & 1.0(3) \\ 
		60 & 0.88474(3) & 1.13028(3) & 194(18) & -2.12(9) \\ 
		61 & 0.88838(8) & 1.12565(10) & 68(18) & 0.2(3) \\ 
		62 & 0.89028(4) & 1.12325(5) & 126(18) & -0.98(14) \\ 
		63 & 0.89222(6) & 1.12080(7) & 93(18) & 1.49(19) \\ 
		64 & 0.89961(11) & 1.11160(14) & 50(18) & 2.9(4) \\ 
		65 & 0.90121(9) & 1.10962(11) & 60(18) & 1.7(3) \\ 
		66 & 0.903030(14) & 1.107382(17) & 406(18) & -1.08(4) \\ 
		67 & 0.90456(7) & 1.10551(9) & 77(18) & 3.0(2) \\ 
		68 & 0.90726(3) & 1.10222(4) & 164(18) & -0.53(11) \\ 
		69 & 0.91266(7) & 1.09570(8) & 81(18) & -1.2(2) \\ 
		70 & 0.91496(4) & 1.09295(5) & 141(18) & 2.93(13) \\ 
		71 & 0.91773(9) & 1.08965(11) & 64(18) & -2.5(3) \\ 
		72 & 0.92196(4) & 1.08465(4) & 152(18) & -0.78(12) \\ 
		73 & 0.92392(7) & 1.08235(9) & 78(18) & 1.5(2) \\ 
		74 & 0.925665(12) & 1.080304(14) & 471(18) & 1.85(4) \\ 
		75 & 0.92985(10) & 1.07545(11) & 60(18) & 2.7(3) \\ 
		76 & 0.936017(16) & 1.068357(18) & 374(18) & 1.17(5) \\ 
		77 & 0.93950(11) & 1.06439(12) & 55(18) & -1.8(3) \\ 
		78 & 0.94242(3) & 1.06110(4) & 179(18) & 2.37(10) \\ 
		79 & 0.94411(10) & 1.05919(12) & 58(18) & 1.5(3) \\ 
		80 & 0.94617(3) & 1.05690(3) & 218(18) & 1.55(8) \\ 
		81 & 0.95602(4) & 1.04600(4) & 172(18) & 1.84(10) \\ 
		82 & 0.96142(8) & 1.04013(9) & 75(18) & 0.1(2) \\ 
		83 & 0.96362(4) & 1.03775(4) & 163(18) & -0.99(11) \\ 
		84 & 0.96843(9) & 1.03260(9) & 72(18) & -3.1(2) \\ 
		85 & 0.97091(5) & 1.02996(5) & 131(18) & -1.95(14) \\ 
		86 & 0.97394(6) & 1.02675(6) & 105(18) & -1.68(17) \\ 
		87 & 0.97970(7) & 1.02073(7) & 97(18) & 0.40(19) \\ 
		88 & 0.98425(3) & 1.01600(3) & 227(18) & -1.37(8) \\ 
		89 & 0.98604(6) & 1.01416(6) & 117(18) & 0.34(15) \\ 
		90 & 0.98885(7) & 1.01128(7) & 97(18) & -2.12(19) \\ 
		91 & 0.99120(13) & 1.00888(13) & 53(18) & -1.4(3) \\ 
		92 & 0.99669(8) & 1.00332(8) & 83(18) & 2.4(2) \\ 
		93 & 1.00486(7) & 0.99516(7) & 97(18) & 1.59(18) \\ 
		94 & 1.00941(3) & 0.99068(3) & 228(18) & 1.12(8) \\ 
		95 & 1.02570(12) & 0.97495(11) & 60(18) & 2.2(3) \\ 
		96 & 1.02970(7) & 0.97116(7) & 96(18) & -3.09(19) \\ 
		97 & 1.03401(6) & 0.96711(5) & 130(18) & 2.90(14) \\ 
		98 & 1.03876(12) & 0.96269(12) & 58(18) & 2.4(3) \\ 
		99 & 1.05079(3) & 0.95166(2) & 280(18) & -0.21(6) \\ 
		100 & 1.05778(12) & 0.94538(10) & 65(18) & 1.6(3) \\ 
		101 & 1.06657(4) & 0.93759(4) & 172(18) & -0.34(10) \\ 
		102 & 1.06932(10) & 0.93517(9) & 75(18) & 1.5(2) \\ 
		103 & 1.07748(11) & 0.92809(9) & 72(18) & -2.4(3) \\ 
		104 & 1.09037(14) & 0.91712(12) & 58(18) & -1.6(3) \\ 
		105 & 1.09466(17) & 0.91353(14) & 48(18) & 0.9(4) \\ 
		106 & 1.10423(15) & 0.90561(13) & 53(18) & -3.0(3) \\ 
		107 & 1.11936(10) & 0.89337(8) & 86(18) & 1.7(2) \\ 
		108 & 1.12408(17) & 0.88962(14) & 49(18) & 0.7(4) \\ 
		109 & 1.13356(9) & 0.88217(7) & 95(18) & 0.54(19) \\ 
		110 & 1.14928(14) & 0.87011(10) & 65(18) & 2.4(3) \\ 
		111 & 1.15379(20) & 0.86671(15) & 46(18) & -0.3(4) \\ 
		112 & 1.16225(9) & 0.86040(6) & 105(18) & 0.08(17) \\ 
		113 & 1.18405(18) & 0.84456(13) & 53(18) & -1.4(3) \\ 
		114 & 1.19466(9) & 0.83706(6) & 111(18) & 3.03(16) \\ 
		115 & 1.22018(19) & 0.81955(13) & 54(18) & -0.7(3) \\ 
		116 & 1.23026(5) & 0.81284(3) & 218(18) & -3.07(8) \\ 
		117 & 1.25609(12) & 0.79612(8) & 90(18) & 0.7(2) \\ 
		118 & 1.26285(15) & 0.79186(9) & 74(18) & 0.2(2) \\ 
		119 & 1.3871(2) & 0.72092(13) & 53(18) & -1.1(3) \\ 
		120 & 1.4171(3) & 0.70564(14) & 50(18) & 0.3(4) \\ 
		121 & 1.4801(3) & 0.67562(12) & 56(18) & 2.2(3) \\ 
		122 & 1.53929(9) & 0.64965(4) & 175(18) & -2.69(10) \\ 
		123 & 1.5613(4) & 0.64050(15) & 46(18) & 2.3(4) \\ 
		124 & 1.6203(3) & 0.61717(13) & 54(18) & -0.7(3) \\ 
		125 & 1.6730(3) & 0.59773(10) & 68(18) & -0.3(3) \\ 
		126 & 1.6919(3) & 0.59104(10) & 69(18) & 0.9(3) \\ 
		127 & 1.7623(4) & 0.56743(11) & 60(18) & -0.7(3) \\ 
		128 & 1.7781(4) & 0.56241(13) & 52(18) & 0.0(3) \\ 
		129 & 1.8100(4) & 0.55247(12) & 57(18) & -3.1(3) \\ 
		130 & 2.2193(7) & 0.45059(14) & 47(18) & -0.1(4) \\ 
		131 & 2.3966(8) & 0.41725(13) & 50(18) & -0.6(4) \\ 
		132 & 2.4259(9) & 0.41222(16) & 43(18) & -1.1(4) \\ 
		133 & 2.6346(10) & 0.37957(14) & 49(18) & 1.1(4) \\ 
		134 & 2.7939(2) & 0.35793(3) & 236(18) & 2.00(8) \\ 
		135 & 2.8831(8) & 0.34685(9) & 72(18) & -0.4(3) \\ 
		136 & 2.9022(12) & 0.34457(14) & 49(18) & -0.4(4) \\ 
		137 & 2.93250(16) & 0.341006(19) & 362(18) & -2.48(5) \\ 
		138 & 2.98957(18) & 0.33450(2) & 326(18) & -0.24(6) \\ 
		139 & 3.0942(6) & 0.32318(7) & 101(18) & 2.56(18) \\ 
		140 & 3.1502(4) & 0.31744(4) & 183(18) & -0.60(10) \\ 
		141 & 3.2153(4) & 0.31101(4) & 182(18) & -2.20(10) \\ 
		142 & 3.3806(6) & 0.29581(5) & 126(18) & -1.93(14) \\ 
		143 & 3.4846(5) & 0.28698(5) & 149(18) & -0.66(12) \\ 
		144 & 3.6597(12) & 0.27325(9) & 76(18) & 0.6(2) \\ 
		145 & 3.7703(7) & 0.26523(5) & 144(18) & 0.88(13) \\ 
		146 & 4.0088(2) & 0.249450(15) & 465(18) & 0.92(4) \\ 
		147 & 4.0443(19) & 0.24726(12) & 58(18) & -2.3(3) \\ 
		148 & 4.1365(9) & 0.24175(5) & 127(18) & -1.13(14) \\ 
		149 & 4.4617(9) & 0.22413(5) & 143(18) & -2.87(13) \\ 
		150 & 4.5734(18) & 0.21865(8) & 80(18) & -1.0(2) \\ 
		151 & 4.9828(10) & 0.20069(4) & 171(18) & -1.58(11) \\ 
		152 & 5.174(2) & 0.19328(8) & 82(18) & 2.6(2) \\ 
		153 & 5.5877(2) & 0.178965(7) & 928(18) & 0.770(19) \\ 
		154 & 5.820(5) & 0.17182(16) & 43(18) & 0.9(4) \\ 
		155 & 6.102(2) & 0.16388(6) & 120(18) & -0.85(15) \\ 
		156 & 6.4293(4) & 0.155539(9) & 722(18) & -1.48(2) \\ 
		157 & 6.774(6) & 0.14762(12) & 56(18) & 2.5(3) \\ 
		158 & 7.152(7) & 0.13983(13) & 51(18) & -2.6(4) \\ 
		159 & 7.6059(15) & 0.13148(3) & 268(18) & 0.64(7) \\ 
		160 & 9.145(4) & 0.10934(5) & 147(18) & -2.64(12) \\ 
		161 & 10.776(12) & 0.09280(10) & 68(18) & -0.6(3) \\ 
		162 & 11.5882(18) & 0.086295(13) & 510(18) & -0.25(4) \\ 
		163 & 14.917(16) & 0.06704(7) & 91(18) & 2.08(20) \\ 
		164 & 16.81(4) & 0.05949(13) & 51(18) & 1.2(4) \\ 
		165 & 18.297(14) & 0.05465(4) & 166(18) & -2.79(11) \\ 
		166 & 21.02(4) & 0.04758(9) & 79(18) & -0.3(2) \\ 
		167 & 22.103(12) & 0.04524(2) & 277(18) & -1.74(7) \\ 
		168 & 32.50(15) & 0.03077(14) & 47(18) & 1.8(4) \\ 
		169 & 36.51(18) & 0.02739(13) & 50(18) & 1.5(4) \\ 
		170 & 42.59(4) & 0.02348(2) & 301(18) & -2.78(6) \\ 
		171 & 46.08(9) & 0.02170(4) & 155(18) & 0.11(12) \\ 
		172 & 52.6(3) & 0.01902(12) & 54(18) & 2.1(3) \\ 
		173 & 78.4(4) & 0.01276(6) & 104(18) & -2.87(17) \\ 
		174 & 154(3) & 0.00649(14) & 50(18) & -1.1(4) \\ 
	\end{longtable}
	%)%end longtab
\end{center}

\clearpage
\twocolumn
		
		\begin{table*}
			\caption{Periods, frequencies, amplitudes, and phases of the modes in the detected prograde dipole mode period series, as extracted by \citet{2021A&A...655A..59V}. Numbers in parenthesis are the errors on the last significant digit.}
			\label{tab:freq_list}
			\centering
			\setlength\tabcolsep{6.0pt} % default value: 6pt
			\renewcommand{\arraystretch}{1.13}
			\newcolumntype{d}[1]{D{.}{.}{#1}}

			\begin{tabular} {r l l r d{3.5}}
				\hline
				\hline
				\# & $p$ [d] & $f$ [d$^{-1}$] & $A$ [ppm] & \multicolumn{1}{c} {$\theta \, [\rm {rad}]$} \\
				\hline
				1 & 0.72834(3)   & 1.37298(5)    & 134(18)  & 2.02(13)  \\
				2 & 0.749329(19) & 1.33453(4)    & 197(18)  & -0.14(9)  \\
				3 & 0.769383(8)  & 1.299742(14)  & 472(18)  & -1.35(4)  \\
				4 & 0.78887(3)   & 1.26764(5)    & 141(18)  & -2.07(12) \\
				5 & 0.8063568(5) & 1.2401457(8)  & 8064(18) & 2.855(2)  \\
				6 & 0.8219273(8) & 1.2166526(12) & 5676(18) & -1.942(3) \\
				7 & 0.8368807(13)& 1.1949135(19) & 3529(18) & -2.774(5) \\
				8 & 0.852339(4)  & 1.173242(6)   & 1197(18) & -1.522(15)\\
				9 & 0.866155(16) & 1.15453(2)    & 320(18)  & 3.04(6)   \\
				10 & 0.87911(6)   & 1.13752(8)    & 82(18)   & -2.5(2)   \\
				11 & 0.89222(6)   & 1.12080(7)    & 92(18)   & 1.49(19)  \\
				12 & 0.903030(14) & 1.107382(17)  & 406(18)  & -1.08(4)  \\
				13 & 0.91496(4)   & 1.09295(5)    & 140(18)  & 2.93(13)  \\
				14 & 0.925665(12) & 1.080304(14)  & 470(18)  & 1.85(4)   \\
				15 & 0.936017(16) & 1.068357(18)  & 374(18)  & 1.17(5)   \\
				16 & 0.94617(3)   & 1.05690(3)    & 218(18)  & 1.55(8)   \\
				17 & 0.95602(4)   & 1.04600(4)    & 172(18)  & 1.84(10)  \\
				18 & 0.96362(4)   & 1.03775(4)    & 163(18)  & -0.99(11) \\
				19 & 0.97091(5)   & 1.02996(5)    & 131(18)  & -1.95(14) \\
				20 & 0.97970(7)   & 1.02073(7)    & 96(18)   & 0.40(19)  \\
				21 & 0.98885(7)   & 1.01128(7)    & 96(18)   & -2.12(19) \\
				22 & 0.99669(8)   & 1.00332(8)    & 83(18)   & 2.4(2)    \\
				\hline
			\end{tabular}
		\end{table*}

		\begin{table*}
			\caption{Periods, frequencies, amplitudes, and phases of the modes in the possible additional period series, as extracted by \citet{2021A&A...655A..59V}. Numbers in parenthesis are the errors on the last significant digit.}
			\label{tab:freq_list_additional}
			\centering
			\setlength\tabcolsep{6.0pt} % default value: 6pt
			\renewcommand{\arraystretch}{1.13}
			\newcolumntype{d}[1]{D{.}{.}{#1}}

			\begin{tabular} {r l l r d{3.5}}
				\hline
				\hline
				\# & $p$ [d] & $f$ [d$^{-1}$] & $A$ [ppm] & \multicolumn{1}{c} {$\theta \, [\rm {rad}]$} \\
				\hline
				1 & 2.93250(16)	& 0.341006(18) & 362(18) & -2.48(5) \\
				2 & 3.1502(4)	& 0.31744(4)   & 183(18) & -0.60(10) \\
				3 & 3.3806(6)	& 0.29581(5)   & 125(18) & -1.93(14) \\
				4 & 3.6597(12)	& 0.27325(9)   & 76(18)  &  0.6(2) \\
				5 & 4.0088(2)	& 0.249450(15) & 465(18) &  0.92(4) \\
				6 & 4.4617(9)	& 0.22413(5)   & 142(18) & -2.87(13) \\
				7 & 4.9828(10)	& 0.20069(4)   & 171(18) & -1.58(11) \\
				8 & 5.5877(2)	& 0.178965(7)  & 928(18) &  0.770(19) \\
				9 & 6.4293(4)	& 0.155539(9)  & 722(18) & -1.48(2) \\
				10 & 7.6059(15)	& 0.13148(3)   & 267(18) &  0.64(7) \\
				11 & 9.145(4)	& 0.10934(5)   & 146(18) & -2.64(12) \\
				12 & 11.5882(18)& 0.086295(13) & 509(18) & -0.25(4) \\
				\hline
				1 & 2.98957(18)	& 0.33450(2) & 326(18) & -0.24(6) \\
				2 & 3.2153(4)	& 0.31101(4) & 181(18) & -2.20(10) \\
				3 & 3.4846(5)	& 0.28698(5) & 149(18) & -0.66(12) \\
				4 & 3.7703(7)	& 0.26523(5) & 143(18) &  0.88(13) \\
				5 & 4.1365(9)	& 0.24175(5) & 127(18) & -1.13(14) \\
				6 & 4.5734(17) 	& 0.21865(8) & 79(18)  & -1.0(2) \\
				7 & 5.174(2)	& 0.19328(8) & 81(18)  &  2.6(2) \\
				\hline
			\end{tabular}
		\end{table*}

	\end{appendix}
	%%%%%%%%%%%%%%%%%%%%%%%%%%%%%%%%%%%%%%%%%%%%%%%%%%%%%%%%%%%%%%%%%%%%%%%%%%%%%%%%%%%%%%%%%%%%%%
\end{document}